\documentclass[11pt,a4paper]{article}
\pdfoutput=1
\usepackage{graphicx}
\usepackage{color}
\usepackage{jcappub}
\usepackage{subfig}
\usepackage{comment}
\usepackage{amsmath} 
   \usepackage{multirow}

\newcommand{\be}{\begin{equation}}
\newcommand{\ee}{\end{equation}}
\newcommand{\bea}{\begin{eqnarray}}
\newcommand{\eea}{\end{eqnarray}}

\newcommand{\gsim}{\lower.7ex\hbox{$\;\stackrel{\textstyle>}{\sim}\;$}}
\newcommand{\lsim}{\lower.7ex\hbox{$\;\stackrel{\textstyle<}{\sim}\;$}}

\usepackage[style=numeric-comp,sorting=none,backend=bibtex]{biblatex}
\title{
Inelastic dark matter scattering off Thallium cannot save DAMA 
}

\author[a]{Sunniva Jacobsen,}

\author[a,b,c]{Katherine Freese,}

\author[d]{Chris Kelso,}

\author[e]{Pearl Sandick,}

\author[a,f,g,h]{Patrick Stengel}

\affiliation[a]{The Oskar Klein Centre for Cosmoparticle Physics,
	Department of Physics,
	Stockholm University,
	AlbaNova,
	10691 Stockholm,
	Sweden}
\affiliation[b]{Nordita,
	KTH Royal Institute of Technology and Stockholm University
	Roslagstullsbacken 23,
	10691 Stockholm,
	Sweden}
\affiliation[c]{
    Department of Physics,
  University of Texas,
  Austin, TX 78722}
\affiliation[d]{Department of Physics, University of North Florida, Jacksonville, FL  32224, USA}
\affiliation[e]{Department of Physics and Astronomy, University of Utah, Salt Lake City, UT 84102, USA}
\affiliation[f]{Scuola Internazionale Superiore di Studi Avanzati (SISSA), via Bonomea 265, 34136 Trieste, Italy}
\affiliation[g]{INFN, Sezione di Trieste, via Valerio 2, 34127 Trieste, Italy}
\affiliation[h]{Institute for Fundamental Physics of the Universe (IFPU), via Beirut 2, 34151 Trieste, Italy}

\abstract{
We study the compatibility of the observed DAMA modulation signal with inelastic scattering of dark matter (DM) off of the $0.1\%$ Thallium (Tl) dopant in DAMA. In this work we test whether there exist regions of parameter space where the Tl interpretation gives a good fit to the most recent data from DAMA, and whether these regions are compatible with the latest constraints from other direct detection experiments. Previously, Chang et al. in 2010~\cite{Chang:2010pr}, had proposed the Tl interpretation of the DAMA data, and more recently (in 2019)  the DAMA/LIBRA collaboration~\cite{Bernabei_2019Modeldependentanalysis} found regions in parameter space of Tl inelastic scattering that differ by more than $10\sigma$ from a no modulation hypothesis. We have expanded upon their work by testing whether the regions of parameter space where inelastic DM-Tl scattering gives a good fit to the most recent DAMA data survive the constraints placed by the lack of a DM signal in XENON1T and CRESST-II. In addition, we have tested how these regions change with the main sources of uncertainty: the Tl quenching factor, which has never been measured directly, and the astrophysical uncertainties in the DM distribution. We conclude that inelastic DM scattering off Tl cannot explain the DAMA data in light of null results from other experiments.
}

\notoc
\makeatletter
\gdef\@fpheader{\begin{flushright}
NORDITA-2020-123; UTTG-24-2020
\end{flushright}}
\makeatother
\begin{document}

\maketitle

\section{Introduction}

The nature of dark matter (DM) is one of the longest outstanding problems of modern physics. Today there is evidence of the gravitational effects of DM on length scales ranging from dwarf galaxies to the largest observable structures of the Universe. One of the most promising candidates for DM is weakly interacting massive particles (WIMPs). In addition to feeling the effects of gravity, these particles can interact weakly, which makes it possible to detect their interactions with ordinary matter. Despite the massive experimental efforts that have been made to directly detect signatures of WIMPs recoiling off of nuclei, no conclusive evidence of DM interactions with ordinary matter has been found~\cite{Agnese:2014SuperCDMS,Akerib:2016LUX,Aprile:2017Xenon1T,Aprile:2018Xenon1TDMsearch,Angloher:2015ewa, Angloher:2014myn, Abdelhameed_2019CRESST3}. 

However, there is one exception to the null-results of direct detection experiments: For the last 20 years the DAMA collaboration has claimed a positive signal of DM scattering, with an accumulated significance of $12.9  \sigma $~\cite{Bernabei:2008DAMAfirstresults, Bernabei:2010newresults, Bernabei:2018phase2} after three phases of data-taking. The DAMA/LIBRA (previously, DAMA/NaI) experiment consists of 250 kg of NaI(Tl) scintillators and is located deep underground at the Gran Sasso National Laboratory. Unlike other direct detection searches in which experiments must be virtually background free to detect a few signal events, DAMA takes advantage of the expected time variation of the DM-induced recoil rate in a detector due to the Earth's motion around the Sun~\cite{Freese:1987wu, Freese:2012xd}. The effect of annual modulation in the WIMP recoil rate was first proposed in the 1980's by Drukier, Freese and Spergel~\cite{1986DrukierFreeseSpergel}. While the likely backgrounds that can mimic the DAMA modulation signal have all been rejected by the collaboration, conventional models for WIMP DM also predict a significant signal from elastic scattering off nuclei in other direct detection experiments, making their null results in strong tension with this interpretation of the DAMA signal, see e.g. Refs.~\cite{Savage_2009, Savage_2009galrotvel, Savage_2011}. DAMA/LIBRA has now published data down to a lower threshold of 1keV~\cite{Bernabei:2018phase2}.  With this new data, Ref.~\cite{Baum:2018ekm} showed that the results of the DAMA/LIBRA experiment are no longer self-consistent under the assumption of vanilla spin independent (isospin conserving) elastic scattering. In addition, recent results from the ANAIS-112 experiment~\cite{Amare:2021yyu}, with the same NaI(Tl) target as DAMA, show a signal consistent with no modulation and the modulation amplitude is in $\sim 3 \sigma$ tension with the signal observed at DAMA.

Thus, it is reasonable to ask if there exist models of DM interactions beyond the standard lore that can explain both the signal in DAMA and the lack of one in all other direct detection experiments. Several alternatives have been proposed, such as DM scattering within a generalized effective field theory framework~\cite{Kang:2018qvz,Kang:2019dbr}, models of mirror DM~\cite{Foot:2008MirrorDM,Adams:2020ejt} and inelastic DM~\cite{TuckerSmith:2001iDM, TuckerSmith:2004iDM}. Inelastic DM (iDM) refers to dark matter that scatters off of target nuclei in direct detection experiments by transitioning to an additional state with a different mass. iDM occurs naturally in several theoretical frameworks, such as certain supersymmetric models~\cite{Fox_2014iDMSUSY}, magnetic iDM~\cite{Chang_2010MiDM}, and dark photon mediated DM~\cite{Smolinsky_2017DarkPhotoniDM}. A recent study of inelastic DM scattering off Na and I at DAMA explores the sensitivity of future annual modulation experiments to several effective models of DM-nucleon interactions which reduce the tension of the observed DAMA signal with null results from other direct detection experiments~\cite{Zurowski:2020dxe}.

In work from 2010, Chang, Lang, and Weiner~\cite{Chang:2010pr} proposed that the $0.1\%$ of thallium (Tl) dopant in the DAMA crystals might be responsible for the annual modulation observed in DAMA. The stable Tl isotopes have atomic mass of $A= 203$ and $A = 205$, and are thus much heavier than most target nuclei in other direct detection experiments (e.g. for xenon, A = 131). When DM scatters inelastically into a heavier state, there exists parameter space where recoils of a given energy are permitted for the heavier Tl and tungsten (W) targets but kinematically forbidden for the lighter xenon (Xe) target. In particular, the minimum velocity required for DM scattering off of lighter targets (e.g. Xe) can be larger than the maximum DM velocity permitted within the Milky Way halo. Thus, in iDM models where the elastic scattering is suppressed, inelastic scattering off of heavy target nuclei (e.g.~Tl) can provide the leading contribution to the DM-nucleus scattering cross-section. In addition to the suppression of scattering off of lighter target nuclei in iDM models, the nuclear form factor for heavy target nuclei such as Tl can provide for the characteristic shape of the nuclear recoil spectrum observed in the DAMA signal. Based on data available at the time, the analysis in Ref.~\cite{Chang:2010pr} suggested that there exist regions of iDM parameter space which could provide a good fit to the DAMA signal and evade constraints from other direct detection experiments. Also, after the most recent phase of data-taking with higher sensitivity to low-energy events, the DAMA collaboration~\cite{Bernabei_2019Modeldependentanalysis} found regions of parameter space where the expected annual modulation of the signal arising from Tl recoils differs by more than $10\sigma$ from the null hypothesis of no modulation.  This result is only part of the story, however.  A  $10\sigma$ improvement over the null hypothesis is not necessarily impressive if the model is still not a good fit to the data.  It is thus vitally important to ensure that a model comparison to the null hypothesis is also coupled with a goodness-of-fit test. 

In this work, we test whether there exist regions of iDM parameter space where the Tl interpretation gives a good fit to the most recent data from DAMA and whether these regions are compatible with constraints from other direct detection experiments, such as XENON1T~\cite{Aprile:2017Xenon1T,Aprile:2018Xenon1TDMsearch} and CRESST-II~\cite{Angloher:2014myn}. 
We also study how the best-fit signal regions depend on the choice of astrophysical parameters and the Tl quenching factor, both of which are subject to large uncertainties. We test the goodness-of-fit for iDM scattering to the observed modulation signal in two cases: Isospin conserving DM interactions with nucleons and an alternative isospin-violating model where Iodine (I) scattering is maximally suppressed in DAMA. We find that there exist regions in the model parameter space that give a good fit to the DAMA modulation but are ruled out by constraints from XENON1T and/or CRESST-II. Thus, iDM scattering off Tl can no longer reconcile the observed annual modulation in DAMA with the lack of a corresponding signal in other direct detection experiments.

The remaining sections of this paper are organized as follows: In Sec.~\ref{sec:InScatt} we give an overview of the calculation of the recoil rate for these dark matter models in the DAMA experiment and discuss the Tl quenching factor. In Sec.~\ref{subsec: analysis strategy} we discuss the details of our analysis, particularly the effects of varying the astrophysical parameters on the modulation signal and the constraints from other direct detection experiments. In Sec.~\ref{sec:BestFit} we present the best-fit regions of these models to the DAMA signal under a variety of assumptions and in Sec.~\ref{sec:Conclusions} we discuss our conclusions.

\section{Inelastic Scattering in DAMA}
\label{sec:InScatt}

Direct detection experiments search for the signatures of nuclear recoils induced by the interactions of DM with an instrumented volume of target nuclei~\cite{Goodman:1984dc}. When elastic scattering of the target nucleus, $X_{\rm SM}$, is either forbidden or highly suppressed in models which contain at least two DM particles $\chi$ and $\chi'$ that are nearly degenerate in mass, inelastic scattering processes of the type $\chi + X_{\rm SM} \to \chi' + X_{\rm SM}$ can provide the leading contribution to the scattering cross section. Such scenarios feature two DM states, $\chi$ and $\chi'$, where $\chi$ is the incident DM particle and $\chi'$ is some heavier state. The mass difference between the states is denoted by $\delta = m_{\chi'}-m_{\chi}$.

The minimum speed required for inelastic scattering to yield a nuclear recoil of energy $E_R$ is then
\begin{equation}
\label{eq: vmin}
    v_{\rm min} = \frac{1}{\sqrt{2 m_X E_R}}\left(\frac{m_X E_R}{\mu} + \delta\right) \ ,
\end{equation}
where $m_X$ is the mass of the target nucleus and $\mu$ is the reduced mass of the DM-nucleus system. In the limit $\delta \to 0$, we recover the usual expression for elastic scattering. As we discuss in more detail below, the minimum speed required for DM with mass $m_\chi \sim 100 \,$GeV to induce nuclear recoils with energies $E_R \simeq \delta \sim {\cal O}(100) \,$keV can exceed the typical escape velocity for the Milky Way halo, $\sim 550 \,$km/s, for target nuclei significantly lighter than Tl. Thus, the inelastic scattering in direct detection experiments with liquid noble gas targets can be significantly suppressed.  Yet, due to the presence of the heavier Tl in the DAMA detector,  in combination with effects of the Tl nuclear form factors on the recoil spectrum observed at DAMA, the DAMA signal can be fit well by Tl recoils.  The goal is to check whether or not the bounds from detectors made of materials lighter than Tl, albeit weakened by kinematic effects, are still strong enough to rule out the hypothesis of inelastic scattering with Tl as an explanation for the DAMA annual modulation data.

The differential recoil rate per unit detector mass for DM-nucleus scattering is
\begin{equation}
\label{eq: differential recoil rate}
     \frac{d R}{d E_{R}}=\frac{2 \rho_{\chi}}{m_{\chi}} \int d^{3} v v f(\vec{v}, t) \frac{d \sigma}{d q^{2}}\left(q^{2}, v\right) \ ,
\end{equation}
where $f(\vec{v},t)$ is the DM velocity distribution function in the detector frame and $d \sigma / d q^{2}$ is the differential scattering cross section with $q^2 = 2 m_X E_{R}$ as the momentum exchange. We take the local DM density to be $\rho_{\chi} = 0.3 \, {\rm GeV/cm^3}$, although different estimates can vary significantly.\footnote{Using global methods for determining the dark matter density in the Milky Way, recent determinations lie in the range $0.3-0.5 \, \rm GeV/cm^3$~\cite{deSalas:2020hbh}. Using the results from local analyses, Salas and Widmark find that the preferred range for most analyses is $\rho_{\chi} = 0.4-0.6 \rm GeV/cm^3$.}  For a target consisting of several types of nuclei the total differential recoil rate is given by
\begin{equation}
    \frac{d R}{d E_{R}} = \sum_{X}\xi_X \left(\frac{d R}{d E_{R}}\right)_X, 
\end{equation}
where $X$ denotes a specific nucleus within the target and $\xi_X$ is the associated mass fraction. The differential cross section for the recoil of a given nucleus is given by
\begin{equation}
\label{eq: diff cross section}
\frac{d \sigma}{d q^{2}}\left(q^{2}, v\right)=\frac{\sigma_{0}}{4 \mu^{2} v^{2}} F^{2}(q) \Theta\left(q_{\max }-q\right) ,
\end{equation}
where $\sigma_{0}$ is the scattering cross section at zero momentum transfer and $F$ is the Helm form factor. The Heaviside step function $\Theta(q_{\rm max}-q)$ comes from the maximal momentum transfer allowed for the case of inelastic scattering, given the upper velocity cutoff set by the escape velocity of the DM.  In order for a nuclear recoil of a given energy to occur, the maximal DM velocity must exceed the minimum velocity to which the detector is sensitive. Thus the Heaviside function can be replaced by imposing a minimum velocity cutoff, $v_{\rm min}$, in the integral of Eq.~\eqref{eq: differential recoil rate}. 

For spin-independent scattering, we can parameterize $\sigma_0$ as:
\begin{equation}
    \sigma_{0}^{\mathrm{SI}}=\frac{4}{\pi} \mu^{2}\left[Z f_{p}+(A-Z) f_{n}\right]^{2} \ ,
\end{equation}
where $Z \, (A-Z)$ is the number of protons (neutrons) in the target and $f_p \, (f_n)$ is the coupling of the WIMP to the proton (neutron). In this analysis we consider two cases:
\begin{subequations}
\begin{alignat}{2}
    & \text{Isospin conserving scattering: } && f_n/f_p = 1  \label{eq: isospin conserving case}\\
    & \text{Isospin violating scattering: } && f_n/f_p = -53/(127-53) \label{eq: isospin violating case}
\end{alignat}
\end{subequations}
The isospin violating case is tuned so that DM-I scattering is maximally suppressed. In this case, the DM only scatters off Tl in DAMA.\footnote{Scattering off Sodium (Na) in DAMA/LIBRA occurs as well, but it is negligible for the DM masses we consider here.} Using the relation $\sigma^{\rm SI}_p = 4\mu_{p}^2 f_p^2/\pi$, the differential scattering rate can be written as
\begin{equation}
\label{eq: diff recoil rate, with velocity integral}
    \frac{d R}{d E_R} = \frac{\rho_{\chi}}{2 m_{\chi}}\frac{\sigma_p^{\rm SI}}{\mu_p^2}\left[Z + (A - Z)\frac{f_n}{f_p}\right]^2 F^2(q)\int_{v> v_{\rm min}}d^3 v \frac{f(\vec{v}, t)}{v} \ ,
\end{equation}
where $\sigma_p^{\rm SI}$ is the proton-DM scattering cross section in the zero momentum transfer limit and $\mu_p$ is the proton-DM reduced mass.

\begin{figure*}
    \includegraphics[width=0.49\linewidth]{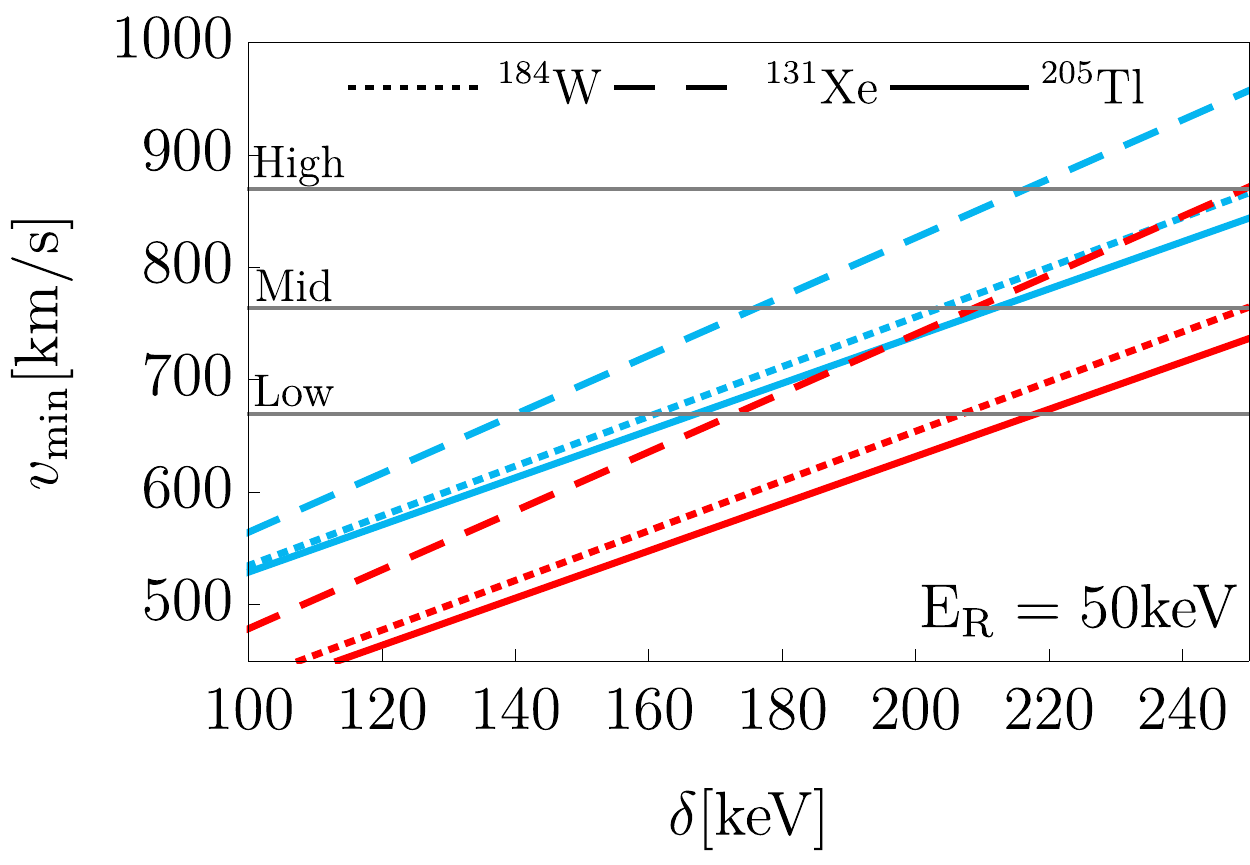}
    \hfill
    \includegraphics[width=0.49\linewidth]{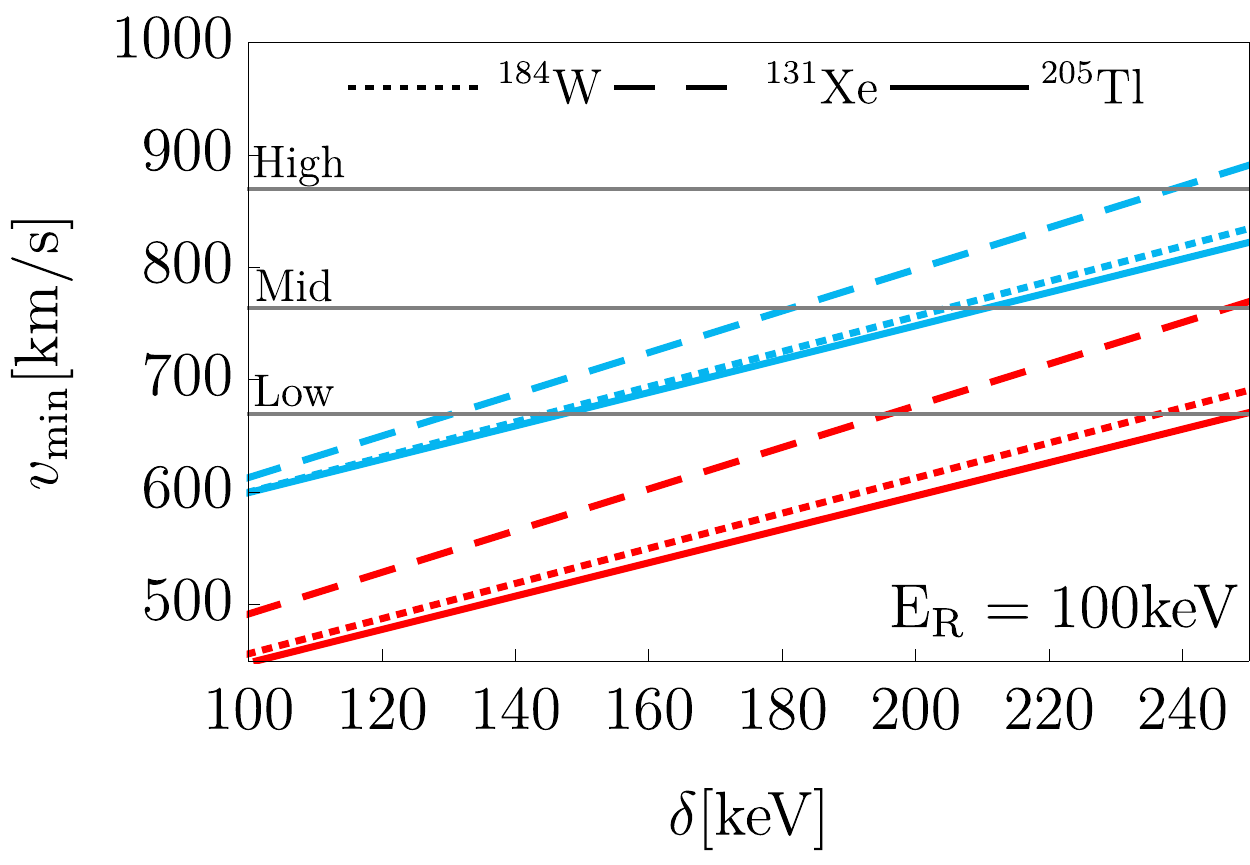}
    \caption{Minimum scattering speed, as a function of mass splitting between DM states, required for a $50\,$keV (left) or $100\,$keV (right) recoil of $\rm ^{205}Tl$ (solid), $\rm ^{131}Xe$ (dashed) or $\rm ^{184}W$ (dotted) induced by an incoming DM state with $m_\chi = 100\,$GeV (blue) or $m_\chi = 200\,$GeV (red). The horizontal grey lines indicate the maximum speed (in the Earth's reference frame) permitted in the galactic halo for benchmarks with $v_{\rm esc} = 440 \,$km/s (Low), $v_{\rm esc} = 544 \,$km/s (Mid) and $v_{\rm esc} = 640 \,$km/s (High), as defined in Eq.~\eqref{eq: high, mid, low}. For our iDM model, we can see that there exists parameter space in the $(m_\chi,\delta)$-plane where recoils of a given energy are permitted for the heavier Tl and W targets but kinematically forbidden for the lighter Xe target.}
    \label{fig:vminvsdelta}
\end{figure*}

As discussed above, the inelastic scattering process we consider can be arranged to favor heavier targets. As shown in Eq~\ref{eq: diff recoil rate, with velocity integral} the differential recoil rate is proportional to the proton and nucleon number squared, and the velocity threshold for scattering, $v_{\rm min}$, decreases with the nucleus mass. For the Tl interpretation of the DAMA signal, the most important effect is the latter. Since Tl is so much heavier than Xe and I, there will be combinations of $m_{\chi}$ and $\delta$ that give a minimum velocity threshold so large that Xe and I scattering cannot occur, while Tl scattering is still allowed. To illustrate this point, in Fig.~\ref{fig:vminvsdelta} we show $v_{\rm min}$ required for 50 keV (left) and 100 keV (right) recoils of W (dotted), Xe (dashed), and Tl (solid), for two different values of $m_\chi$, as a function of $\delta$.  We see that there exists parameter space in the $(m_\chi, \delta)$-plane where recoils of a given energy are permitted for the heavier Tl and W targets but kinematically forbidden for the lighter Xe target. Also, the minimum velocity generally increases as the DM mass decreases. Thus, heavier atoms such as Tl will be able to scatter with lighter DM that is kinematically unavailable to lighter atoms like Xe. Note that I and Xe are kinematically similar, and thus the minimum velocity required for DM-I scattering will have the same dependence on the mass splitting as Xe. 

To find the differential recoil rate in the detector, we must take into account the finite energy resolution of the experiment. The differential recoil rate in the detector can be re-written in terms of the electron equivalent of the nuclear recoil energy, $E_{ee}$, as
\begin{equation}
\label{eq:dRdEee}
    \frac{d R}{d E_{ee}} =\int_{0}^{\infty} d E_R \phi\left(E_R, E_{ee}\right) \frac{d R}{d E_R}\ ,
\end{equation}
where $\phi (\rm E_{R}, E_{ee})$ is the differential response function. Since the DAMA collaboration presents their results corrected for the efficiency of the detector, we have omitted any such correction in our calculation of the recoil rate. The response function is given by
\begin{equation}
\label{eq:res_func}
    \phi\left(E_R, E_{ee}\right)=\frac{1}{2 \pi \sigma_E^{2}} \exp{ \left( -  \frac{(E_{ee} - Q_X E_R)^2 }{ 2 \sigma_E^2} \right)} ,
\end{equation}
where $\sigma_E (Q_X, E_R)$ is the energy resolution of the detector and $Q_X$ is the quenching factor for a given target nucleus. We take the energy resolution to be
\begin{equation}
    \sigma\left(Q_X, E_R\right)=\alpha \sqrt{Q_X E_R}+\beta Q_X E_R \ ,
\end{equation}
with $\alpha=(0.448 \pm 0.035) \sqrt{\mathrm{keV}_{\mathrm{ee}}}$ and $\beta =(9.1 \pm 5.1) \times 10^{-3}$ \cite{Bernabei:2008apparatus}. We use the central values in our calculations and note that changing the resolution has little effect on our conclusions since it is smaller than the size of the energy bins we use after rebinning (see Table~\ref{tab:rebinned modulation amplitudes}). However, changing the quenching factor of the detector nuclei has a significant effect on our signal because it causes the recoils to shift from one energy to another.

\subsection{Quenching Factor}

When a DM particle induces a nuclear recoil in a detector, the energy from the recoil is transferred either to electrons or other nuclei. If the energy is transferred to electrons, the recoil can be observed as scintillation light or ionization. If it is transferred to nuclei, the recoil can be observed as phonons or heat. An experiment that measures scintillation light or ionization usually reports the recoil energy in terms of the electron-equivalent energy, $E_{ee}$, the amount of energy required for an electron recoil to produce the same amount of scintillation light as a nuclear recoil. Generally, nuclear recoils produce less scintillation light than electronic recoils. The true recoil energy and the measured energy are connected by the quenching factor, $E_{ee} = Q_X E_{{R}}$, which is the most likely value of the measured energy in the distribution corresponding to the response function given by Eq.~\eqref{eq:res_func}. The quenching factor is unique to each nucleus and generally depends on the recoil energy. However, here we assume that the quenching factor is constant as a function of energy for a given type of detector nucleus within the energy range relevant for DAMA. 

The Tl quenching factor has never been measured directly. Chang et al.~\cite{Chang:2010pr}  found an approximate range by calculating path lengths and assuming an inverse proportionality to mass. This gives the following relation between the Tl and I quenching factors: $0.62 < Q_{ \rm Tl}/Q_{\rm I} < 0.88$. Using the range of I quenching factors, $0.06< Q_{ \rm I}<0.09$, we get that $0.037<  Q_{\rm Tl}< 0.079$. Given the lack of a direct measurement, we calculate the best-fit regions for Tl scattering in DAMA with $ Q_{\rm Tl}$ ranging from $0.03$ to $0.09$ with a step of $0.01$. As we will demonstrate below, only $ Q_{\rm Tl} = 0.04$ yields a sizeable 2$\sigma$ best-fit region to the DAMA signal. Thus, we assume this value as a benchmark when investigating the effects of varying other parameters. 

\begin{figure}
\centering
    \includegraphics[scale=0.6]{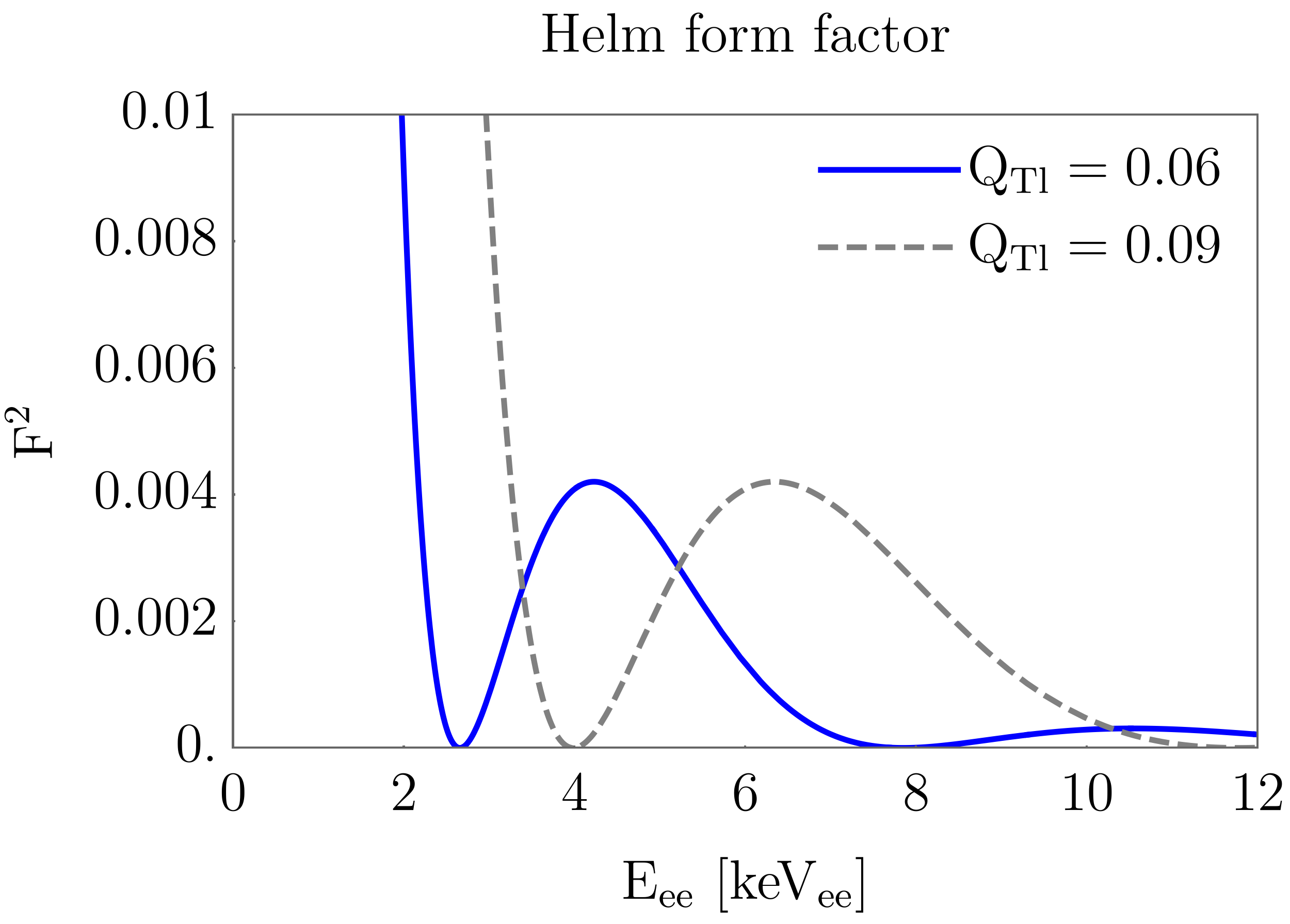}
    \includegraphics[scale=0.6]{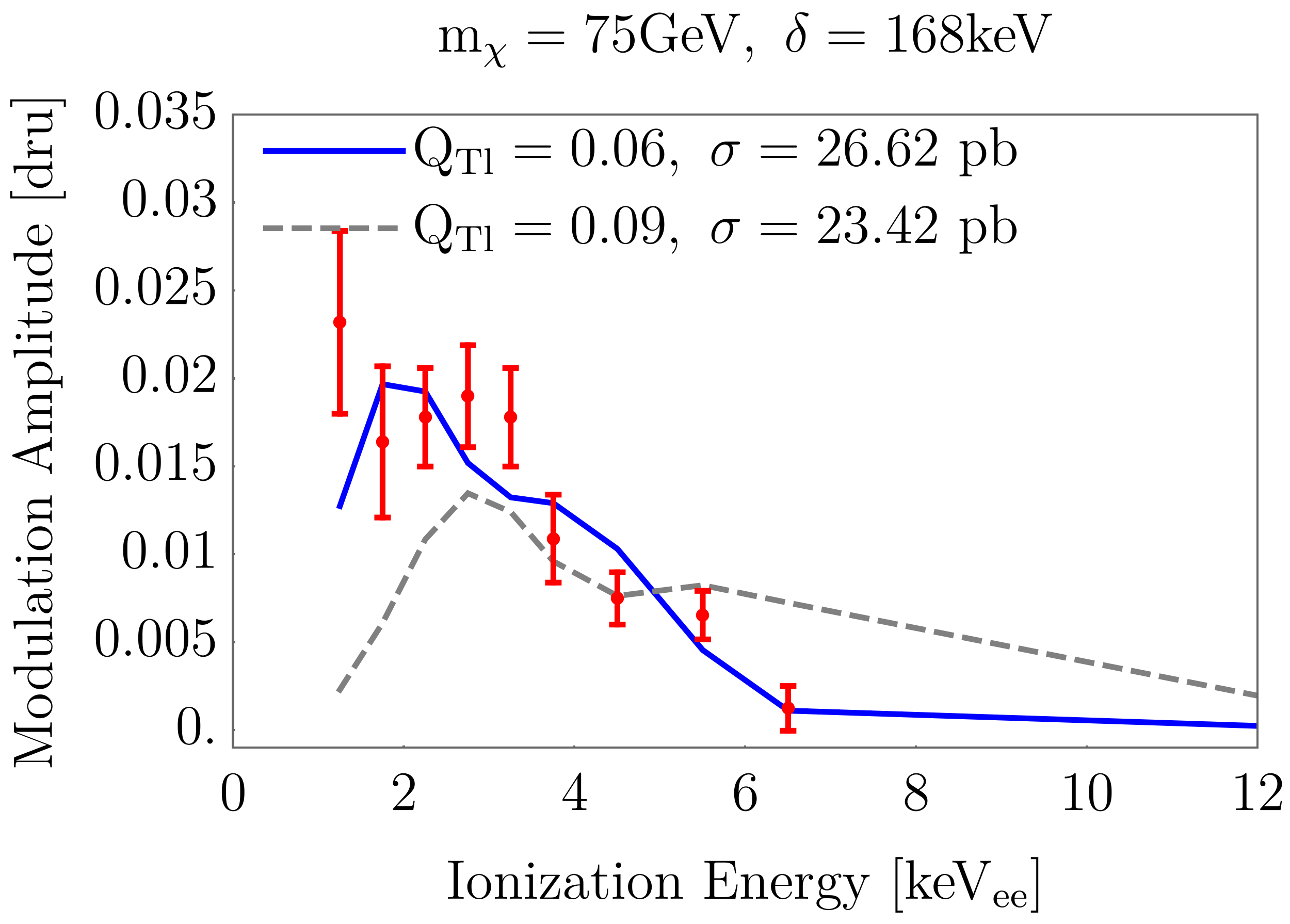}
    \caption{\emph{Left:} The Helm form factor squared for Tl, in the DAMA energy range. \emph{Right:} The expected Tl signal in the isospin-conserving case for two different quenching factors using the best-fit cross sections. Note that the combination of $m_{\chi}$ and $\delta$ give a minimum velocity too large for DM-I scattering to occur. The red points with error bars are the DAMA signal in each energy bin. Increasing the quenching factor of Tl corresponds to shifting the zero-points of the form factor toward larger energies and increasing the period of the oscillatory features in the form factor. This is because a larger Tl quenching factor, $Q_{ \rm Tl}$, for a fixed nuclear recoil energy, $E_{{R}}$, corresponds to a larger electron-equivalent energy, $E_{ee} = Q_{ \rm Tl} E_{{R}}$.}
    \label{fig: form factor and energy spectrum, different QFs}
\end{figure}

In comparison to the recoil spectra of lighter nuclei, Tl recoils are particularly dependent on the quenching factor because of the oscillatory features of the nuclear form factor, $F$, in the energy range relevant to the signal observed at DAMA.  Specifically, the Tl form factor falls to zero in the middle of the DAMA energy range, giving rise to a two-peak shape of the signal spectrum. When the quenching factor increases, the peaks are shifted towards higher energies and become wider. Hence, the Tl quenching factor strongly affects the shape of the expected Tl scattering signal, explaining why the best-fit regions depend so much on the quenching factor. This is illustrated in Fig.~\ref{fig: form factor and energy spectrum, different QFs}, where we can see that the same combination of $m_{\chi}$ and $\delta$ can give two very different signals for two different quenching factors.

\subsection{Annual Modulation of Dark Matter Velocity Distribution}
\label{subsec: annual modulation}

The velocity distribution of the DM particles in the Earth's rest frame (lab frame), $f(\Vec{v}, t)$, changes throughout the year due to the orbital motion of the Earth around the Sun. The velocity distribution in the lab frame is found by performing a Galilean boost of the velocity distribution in the DM rest frame, $\tilde{f}(\Vec{v})$:
\begin{equation}
    f(\Vec{v}, t) = \tilde{f}(\Vec{v}_{\rm lab}(t)+\Vec{v}, t) \ ,
\end{equation}
where $\vec{v}_{\rm lab}(t) = \vec{v}_{\odot} + \vec{v}_{E}(t)$ is the motion of the lab frame relative to the DM rest frame, $\vec{v}_{E}(t)$ is the velocity of the Earth relative to the Sun, $\vec{v}_{\odot} = \vec{v}_{\rm LSR}+ \vec{v}_{\odot,pec}$ is the velocity of the Sun relative to the DM rest frame, $\vec{v}_{\rm LSR} = (0, v_{0},0)$ is the motion of the Local Standard of Rest (in Galactic coordinates) and $\vec{v}_{\odot,pec} = (11,12,7) \,$km/s is the Sun's peculiar velocity. We take the local circular velocity to be $v_0 = 220 \,$km/s.

We assume that the DM velocity distribution can be described by a Maxwell-Boltzmann distribution truncated at the DM escape velocity $v_{\rm esc}$,
\begin{equation}
    f(\vec{v})=\frac{1}{N_{\mathrm{esc}}\left(\pi v_{0}^{2}\right)^{3 / 2}} \exp \left(-\frac{\left|\vec{v}+\vec{v}_{E}\right|^{2}}{v_{0}^{2}}\right) \Theta\left(v_{\mathrm{esc}}-\left|\vec{v}+\vec{v}_{E}\right|\right) \ ,
\end{equation}
where the normalization factor is given by
\begin{equation}
   N_{\mathrm{esc}}=\operatorname{erf}\left(\frac{v_{\mathrm{esc}}}{v_{0}}\right)-\frac{2}{\sqrt{\pi}} \frac{v_{\mathrm{esc}}}{v_{0}} \exp \left(-\frac{v_{\mathrm{esc}}^{2}}{v_{0}^{2}}\right) \ .
\end{equation}
Assuming smooth components of the halo, we can divide the differential recoil rate into two terms: One that includes the constant/unmodulated recoil rate, $S_0$, and a modulation term, $S_m$, that includes the time-variation of the rate due to the Earth's velocity around the Sun:
\begin{equation}
    \frac{d R}{d E}(E, t) \approx S_{0}(E)+S_{m}(E) \cos \omega\left(t-t_{0}\right) \ ,
\end{equation}
where $t_0$ is the date when $v_E$ is maximal. The average modulation amplitude in each energy bin of the DAMA detector is then
\begin{equation}
\label{eq: modulation amplitude}
    \bar{S}_m = \sum_X \frac{\xi_X}{2(E_2-E_1)}\int_{E_1}^{E_2} dE_{ee} \left[\frac{dR}{dE_{ee}}\big(E_{ee}, t_0\big) -\frac{dR}{dE_{ee}}\big(E_{ee}, t_0+0.5\rm \, yrs\big)\right] \ ,
\end{equation}
where $E_1$ and $E_2$ are the lower and upper energy limits of each bin and the recoil rate is given by Eq.~\eqref{eq:dRdEee}. 

The astrophysical parameters $v_0$ and $v_{\rm esc}$, the local circular velocity and the escape velocity of the DM particles, can be subject to significant uncertainties. Typical values used in direct DM detection experiments are $v_{\rm esc} = 544\substack{+64 \\ -46} \,{\rm km/s}$~\cite{Smith_2007RAVEsurvey, Piffl:2013RAVEsurvey} and $v_0 =221 \pm 18 \,$km/s \cite{Koposov_2010}. As mentioned above we take $v_0=220$ km/s and consider three cases for the maximum DM velocity in the Earth's reference frame denoted as: \emph{High}, \emph{Mid}, and \emph{Low}:
\begin{equation}
\label{eq: high, mid, low}
\begin{aligned}
& & v_{\rm esc} &+ v_{\rm lab}(t_0)  \\
\hline
&\text{High:}&640\rm km/s &+ 250 \rm km/s \\
&\text{Mid:} &544\rm km/s &+ 250 \rm km/s \\
&\text{Low:} & 440\rm km/s &+ 250 \rm km/s \\
\end{aligned}
\end{equation}

\begin{figure}

    \includegraphics[width=0.49\linewidth]{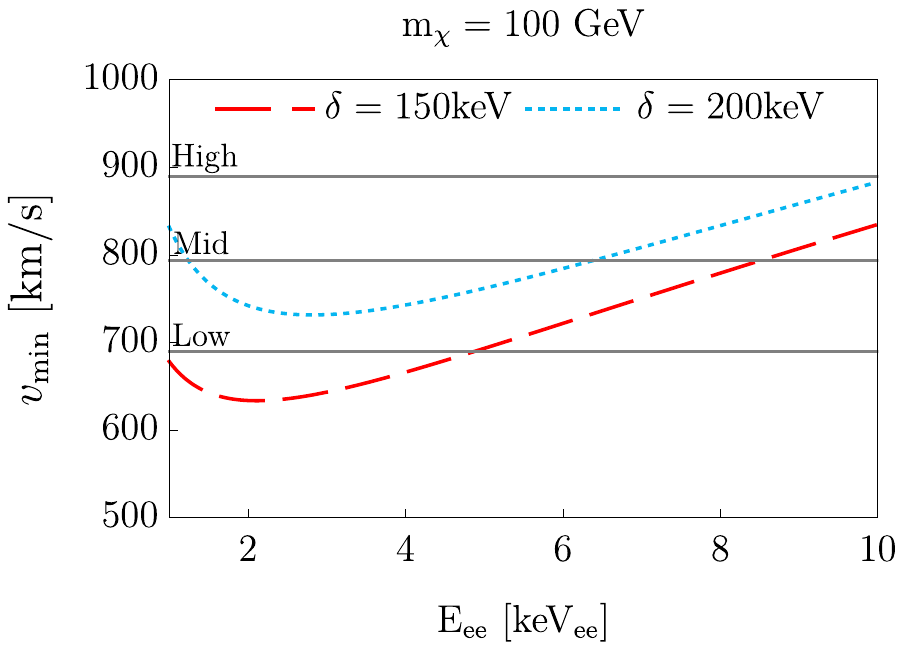}
    \includegraphics[width=0.49\linewidth]{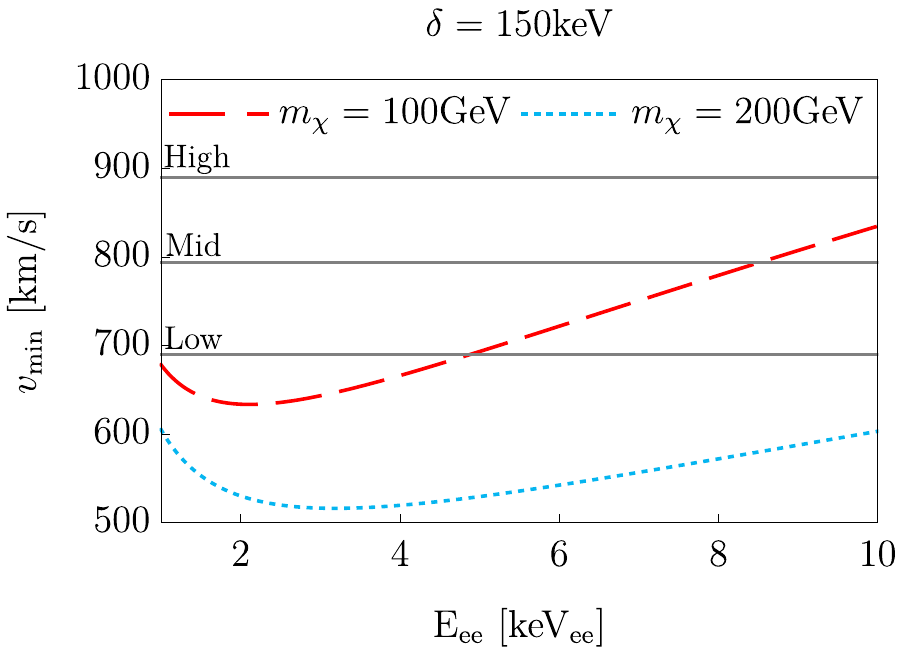}
    \caption{\emph{Left:} The minimum DM velocity required for scattering as a function of the electron equivalent recoil energy for mass splittings $\delta$ = 150 keV (red dashed) and $\delta$ = 200 keV (blue dotted), with the DM mass fixed at $m_\chi = 100 \,$ GeV. \emph{Right}: The minimum velocity as a function of the electron equivalent recoil energy for DM masses $m_\chi = 100 \,$ GeV (red dashed) and $m_\chi = 200 \,$ GeV (blue dotted), with the mass splitting fixed at $\delta$ = 150 keV. In both plots, the minimum velocity is calculated for DM-$^{205}{\rm Tl}$ scattering with a Tl quenching factor of $Q_{\rm Tl }=0.04$. The grey lines give the maximum velocities of the DM particles as seen in the detector for three different cases, defined as in Eq.~\eqref{eq: high, mid, low}. For scattering to occur, the minimum velocity must be smaller than the maximal DM velocity. When $\delta$ increases or $m_{\chi}$ decreases, a larger escape velocity is required in order to maximize the signal in each bin. Thus, both the shape and normalization of the signal depend on the minimum velocity. Note that for clarity in these plots we are not accounting for the finite energy resolution of the detector and taking the central value of the energy distribution given by the response function in Eq.~\eqref{eq:res_func}.}
    \label{fig: minimum velocity with two vesc}
\end{figure}

The minimum velocity for DM-Tl scattering given in Eq.~\ref{eq: vmin} is dominated by the term $\delta / \sqrt{2 m_{X}E_{R}}$ when the recoil energy is relatively small, and otherwise depends approximately linearly on $E_R$ in the parameter space of interest. This leads to the dependence of $v_{\rm min}$ on the (electron equivalent) recoil energy illustrated in Fig.~\ref{fig: minimum velocity with two vesc}, with a minimum somewhere in the low-energy part of the spectrum. Scattering is allowed for a given $v_{\rm esc}$ when the minimum velocity is lower than the horizontal lines corresponding to the maximal DM velocity $v_{\rm esc} + v_{\rm lab}(t)$. Lower escape velocities give smaller intervals of $ E_{\rm ee}$ where scattering is allowed. Because of the shape of $v_{\rm min}$, the corresponding recoil spectrum will have less signal in the low ($E_{\rm ee} < 2 \, {\rm keV}$) and high ($E_{\rm ee} > 7 \, {\rm keV}$) energy bins compared to a recoil spectrum with a higher escape velocity.

\section{Analysis Strategy}
\label{subsec: analysis strategy}

In order to improve the signal-to-noise ratio, an alternative binning of the DAMA signal has been used, according to the procedure in Ref.~\cite{Kelso_2013}. DAMA presents their recoil spectrum in 36 energy bins, each with a width of $0.5$ keV$_{\rm ee}$. Using 36 bins has two main problems: Most of the bins at high recoil energies have a width smaller than the energy resolution of the detector, and a WIMP signal at DAMA's higher energies is typically negligible.  These additional bins thus increase statistical noise without adding information about the signal. The re-binning employed in our analysis groups together adjacent bins with widths substantially more narrow than the energy resolution and groups together all higher-energy bins into one single bin. The resulting bins and their average modulation amplitudes are presented in Table \ref{tab:rebinned modulation amplitudes}. 

\begin{table}[ht!]
    \centering
    \begin{tabular}{c|c} \hline\hline
        Energy & Average $S_m$ \\ 
        $\rm [keV_{ee}]$ & $[\mathrm{cpd} / \mathrm{kg} / \mathrm{keVee}]$ \\ \hline
        $1.0-1.5$ &   $0.0232 \pm 0.0052$ \\
        $1.5-2.0$ &  $0.0164 \pm 0.0043$\\
        $2.0-2.5$ &  $0.0178 \pm 0.0028$ \\ 
        $2.5-3.0$ & $0.0190 \pm 0.0029$  \\
        $3.0-3.5$ &  $0.0178 \pm 0.0028$\\ 
        $3.5-4.0$ &  $0.0109 \pm 0.0025$\\ 
        $4.0-5.0$ &  $0.0075 \pm 0.0015$\\ 
        $5.0-6.0$ &  $0.0066 \pm 0.0014$\\ 
        $6.0-7.0$ &  $0.0013 \pm 0.0013$\\ 
        $7.0-20.0$ &  $0.0007 \pm 0.0004$\\ \hline\hline
    \end{tabular}
    \caption{Average modulation amplitudes observed by DAMA over the given energy bins, after rebinning as in Ref.~\cite{Kelso_2013}. We have used the modulation amplitudes for the whole data sets: DAMA/NaI, DAMA/LIBRA phase-1 and DAMA/LIBRA phase-2, as presented in Ref.~\cite{Bernabei:2018FirstModelIndependentResultsPhase2}. The rebinning is performed in order to improve the signal-to-noise ratio.}
    \label{tab:rebinned modulation amplitudes}
\end{table}
We have performed a goodness-of-fit test with the theoretical modulation amplitude in Eq.~\eqref{eq: modulation amplitude} and the amplitude detected in the DAMA/LIBRA experiment \cite{Bernabei:2018FirstModelIndependentResultsPhase2}. We use a $\chi^2$-test to find the best-fit combinations of $m_{\chi}$, $\sigma_p^{\rm SI}$, and $\delta$,
\begin{equation}
    \chi^2 = \sum_{k=1}^{10}\frac{(S^T_{m,k}(m_{\chi}, \delta, \sigma^{\rm SI}_p) - S_{m,k})^2}{\sigma_k^2} \ ,
\end{equation}
where $S^T_{m,k}(m_{\chi}, \delta, \sigma_p^{\rm SI})$ is the theoretically expected amplitude in bin $k$, $S_{m,k}$ is the measured amplitude and $\sigma_k$ is the corresponding standard error from the DAMA data. We use the rebinned modulation amplitudes and errors presented in Table~\ref{tab:rebinned modulation amplitudes}. For each $m_{\chi}$ and $\delta$, we minimize $\chi^2$ with respect to the cross section $\sigma_p^{\rm SI}$. Note that our confidence regions represent the confidence regions from a perfect fit with the data. The number of degrees of freedom is equal to the number of bins minus the number of free parameters. Using the rebinned data, this gives a total of 7 d.o.f. since we are fitting three parameters: $m_{\chi}, \delta$, and $\sigma^{\rm SI}_p$. We do not include the Tl quenching factor as a free parameter, since we find the confidence regions in $(m_{\chi}, \delta)$-space for fixed values of the quenching factor.

\subsection{Effects of varying astrophysical parameters}

In this section, we discuss how varying astrophysical parameters can change the recoil spectra for iDM scattering in the DAMA detector. We focus on the case where isospin-violation suppresses DM-I scattering in order to isolate the effects of the astrophysical parameters on DM-Tl scattering. We keep the local DM density fixed at $\rho_{\chi} = 0.3 \, \rm GeV/cm^3$ since the associated changes in normalization of recoil spectrum can always be absorbed into the scattering cross section. Note that we only change one parameter at a time, so for all $v_{\rm esc}$ the circular velocity is fixed at the typical value of $v_0 = 220 \,$km/s, and when we vary $v_0$ we keep the escape velocity fixed at $v_{\rm esc} = 544 \,$km/s.

\begin{figure}
    \centering
     \includegraphics[width=0.49\linewidth]{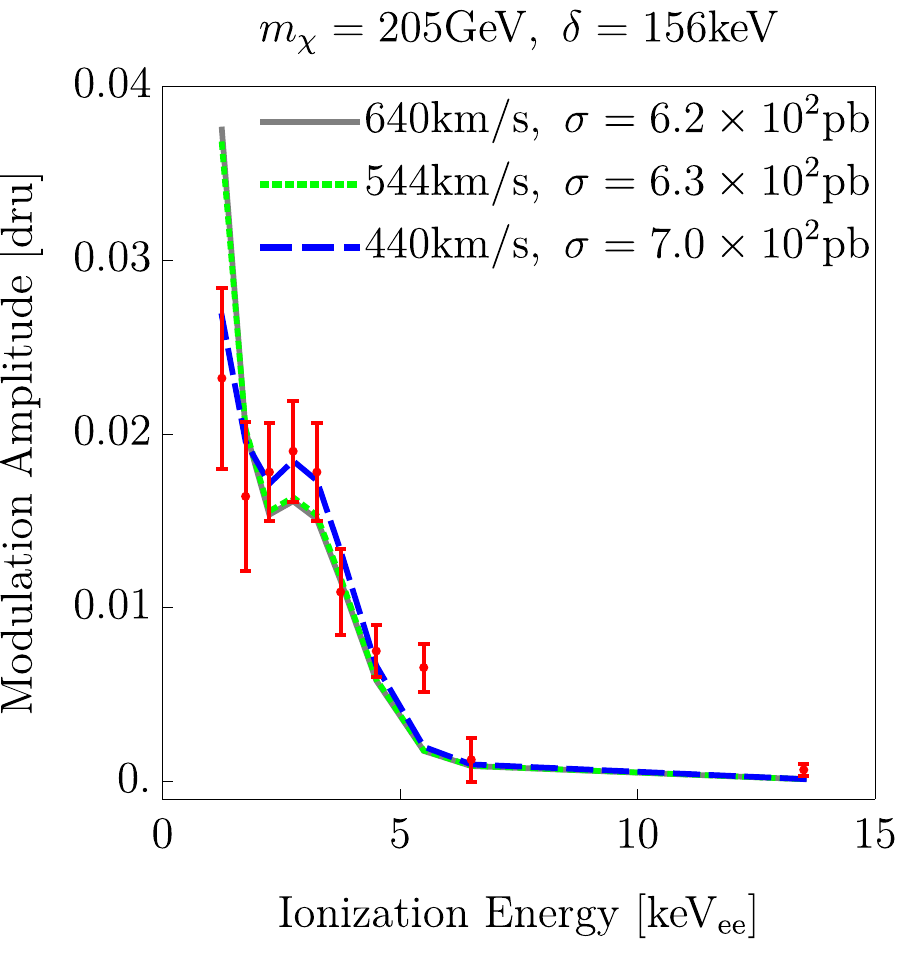}
     \includegraphics[width=0.49\linewidth]{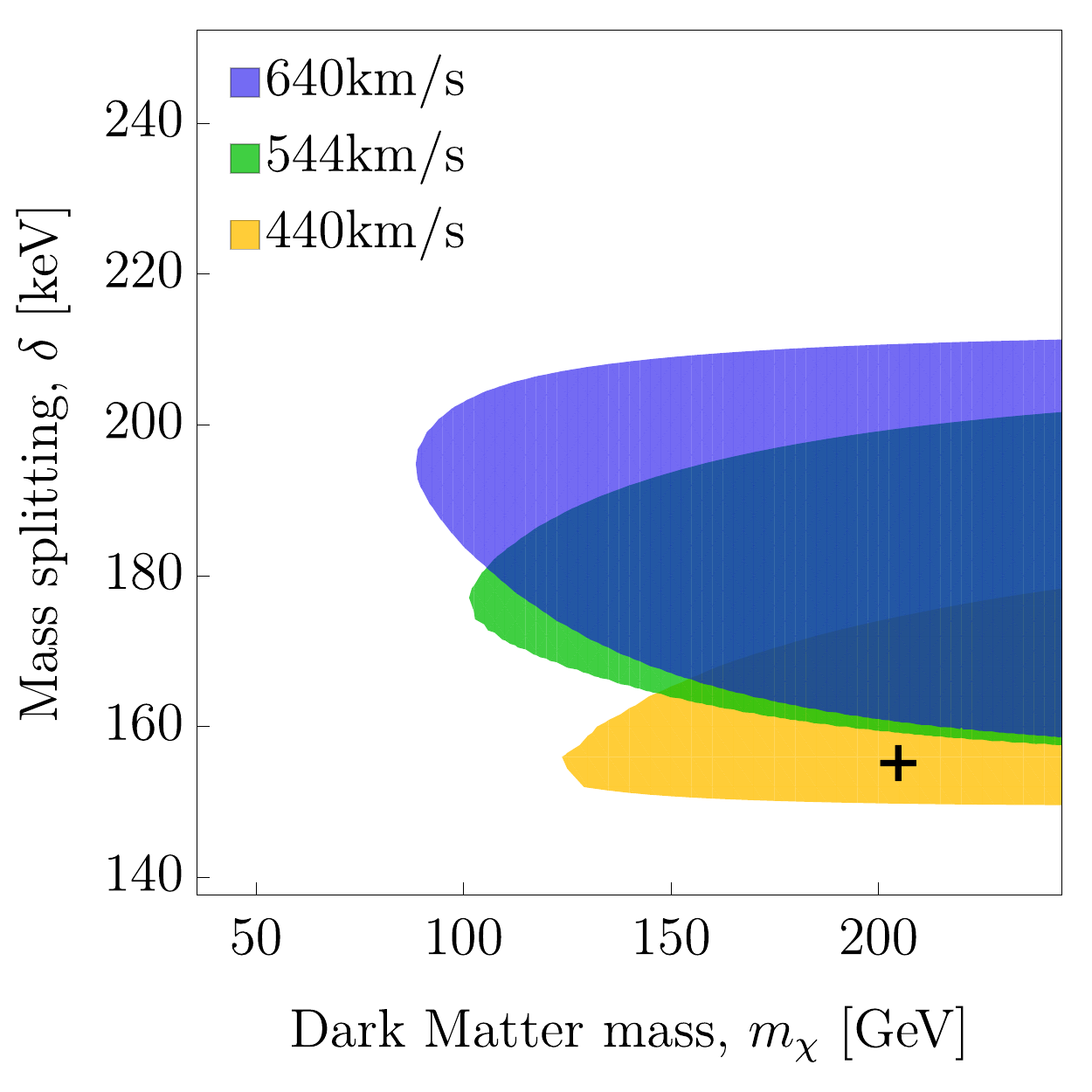}
    \caption{{\it Left}: The expected signal for the best-fit cross section for isospin-violating DM-Tl scattering with $m_{\chi} = 205 \, \rm GeV$ and $\delta = 156 \, \rm keV$ for three different $v_{\rm esc}$.  The red points are the observed signal in DAMA. {\it Right}: The best-fit regions within $3\sigma$ for isospin violating DM-Tl scattering when $Q_{\rm Tl} = 0.04$, for three different DM escape velocities. The best-fit regions increase in size to include higher $\delta$ and smaller $m_{\chi}$ when the escape velocity increases because the DM velocity becomes large enough to allow scattering in regions previously forbidden by the threshold $v_{\rm esc} + v_{\rm lab}(t) > v_{\rm min}$, where $v_{\min}$ is given by Eq.~\eqref{eq: vmin}. The marked point in the best-fit region is located at $m_{\chi} = 205 \, \rm GeV$ and $\delta = 156 \, \rm keV$, showing that this combination of DM mass and mass splitting gives a good fit to the data when $v_{\rm esc} = 440 \, {\rm km/s}$, but not when $v_{\rm esc}$ increases.}
    \label{fig: bf regions, vesc}
\end{figure}

As an examination of how well the DAMA signal can be fit in our model, we study the dependence of the best-fit regions on the escape velocity. As shown in the left panel of Fig.~\ref{fig: bf regions, vesc}, the DAMA signal peaks at lower energies and then falls off, with almost no signal and very small errors in the high energy bins. When $v_{\rm esc}$ increases, more signal is allowed in the high energy bins.  Since the error is so small in the high energy bins, the fitting procedure will prioritize these data points and choose smaller $\sigma_p^{\rm SI}$ so that the overall normalization of the signal is reduced in order to achieve the best fit. In turn, the smaller normalization yields a worse fit to the low-energy part of the spectrum, where the DAMA signal is largest. As $v_{\rm esc}$ increases, this fit will eventually be so bad that it is no longer within $3\sigma$. The right panel of Fig.~\ref{fig: bf regions, vesc} shows an example of this: The combination of $m_{\chi} = 205 \,$GeV and $\delta = 156 \,$keV gives a fit within $3\sigma$ when $v_{\rm esc} = 440 \,$km/s, but not for higher $v_{\rm esc}$. 

As one can see from the figure, the fits with higher escape velocities become worse in the low energy bins of the DAMA signal (see the dotted and solid lines corresponding to $v_{\rm esc} = 544, 640 \,$km/s), while maintaining a good fit to the high-energy bins. Recalling the earlier discussion of how changes in escape velocity can impact the recoil kinematics for various choices of $m_\chi$ and $\delta$, we can then analyse the associated changes to the best-fit regions in the $(m_\chi,\delta)$ plane in the right panel of Fig.~\ref{fig: bf regions, vesc}. When the escape velocity increases, the best-fit regions within $3\sigma$ increase in size and are shifted towards higher $\delta$. When $v_{\rm esc}$ increases, more regions of parameter space open up for scattering because the threshold, $v_{\rm min} < v_{\rm esc} + v_{\rm lab}$, allows for higher $v_{\rm min}$. Since the minimum velocity increases with $\delta$ and decreases with $m_{\chi}$, the best-fit regions expand in size towards higher $\delta$ and lower DM masses. Due to the increasingly poor fit to the low energy part of the DAMA signal as $v_{\rm esc}$ continues to increase, the best-fit regions also shift towards higher $\delta$ and lower $m_\chi$ when $v_{\rm esc}$ increases.

Similar to $v_{\rm esc}$, when the circular velocity increases the best-fit regions are expanded to include larger $\delta$ and smaller $m_{\chi}$. Since $v_{\rm lab} \sim v_{E}(t) + v_0$, the maximum DM velocity increases when $v_0$ increases so that more combinations of $\delta$ and $m_{\chi}$ are allowed.  But, since the uncertainties in $v_0$ are smaller relative to the maximum DM velocity in the Earth's reference frame, this effect is less significant than the corresponding effect for $v_{\rm esc}$. Also, when $v_0$ increases the best-fit regions are shifted towards higher $\delta$. This is due to the velocity integral in Eq.~\eqref{eq: diff recoil rate, with velocity integral}. When $v_0$ increases the velocity integral, or mean inverse speed, is amplified, but not by the same amount for all energies; the velocity integral is generally more amplified in the high-energy bins than the lower energy bins. This leads to a similar effect as for $v_{\rm esc}$: when $v_0$ increases we have too much signal in the high energy bins to find a fit within $3\sigma$ to the DAMA data. 

\subsection{Constraints from other direct detection experiments}

The lack of a DM signal in all direct detection experiments other than DAMA puts significant constraints on DM models. In this work we have chosen to focus on XENON1T~\cite{Aprile:2017Xenon1T, Aprile:2018Xenon1TDMsearch} and CRESST-II~\cite{Angloher:2014myn} to constrain our best-fit models for DM-Tl inelastic scattering in DAMA. We will show that the lack of a DM signal in these experiments rules out all of the parameter space relevant for DM-Tl inelastic scattering.  In this section we present how we have calculated the constraints from XENON1T and CRESST-II, and the effects that varying the astrophysical parameters have on these constraints. 

In order to find the regions of $(m_{\chi}, \delta)$ space where the Tl interpretation of the DAMA signal is ruled out by XENON1T and/or CRESST-II, we calculate the expected number of events for a set of parameters where the Tl interpretation gives a good fit to the DAMA data. The number of expected events is,
\begin{equation}
    N(m_{\chi}, \delta, \sigma_p^{BF}) = MT\sum_X \xi_X\int_{E_{\rm min}}^{E_{\rm max}} dE_{ee} \left[\frac{dR}{dE_{ee}}\big(E_{ee}, t_0\big) +\frac{dR}{dE_{ee}}\big(E_{ee}, t_0+0.5\rm \, yrs\big)\right] \ ,
\end{equation}
where $\frac{dR}{dE_{ee}}\big(E_{ee}, t \big) = \frac{dR}{dE_{ee}}\big(E_{ee}, t, m_{\chi}, \delta, \sigma_p^{BF}\big)$, $MT$ is the exposure of the detector ($M$ is the total target mass and $T$ is the exposure time) and $E_{\rm min}$ and $E_{\rm max}$ are the minimum and maximum recoil energies we consider for a given experiment. For each $(m_{\chi}, \delta)$ in the parameter space presented in Figs.~\ref{fig: 2sigma best-fit regions, isospin-violating} and~\ref{fig: BF regions 3sigma} we find the best-fit cross section for DM-Tl scattering in DAMA, $\sigma_p^{BF}$, and calculate the number of expected events in XENON1T and CRESST-II with this cross section. Since the DAMA crystals only contain $0.1\%$ of Tl dopant, the cross sections required to get a good fit to the DAMA data are typically $\sim 1000$ times larger than what would be required for DM-I scattering to give a good fit. Using this approach, we find that all the best-fit regions presented in Figs.~\ref{fig: 2sigma best-fit regions, isospin-violating} and~\ref{fig: BF regions 3sigma} are ruled out by a combination of XENON1T and CRESST-II, as they predict a number of events in these experiments that far exceed what has actually been observed.

Below we describe constraints for the isospin conserving and isospin violating benchmarks defined by Eqs.~\ref{eq: isospin conserving case} and~\ref{eq: isospin violating case}. Note we have also considered isospin violating cases tuned such that DM scattering off either Xe or W is maximally suppressed in order relax the respective constraints from XENON1T or CRESST-II. When DM-W scattering is maximally suppressed, the associated regions of best fit to the DAMA signal are similar to those of isospin conserving case described in Sec.~\ref{sec:BestFit}. When DM-Xe scattering is maximally suppressed, the best fit regions are also similar to those of the isospin conserving case, but the cutoff where DM-I scattering no longer affects the overall signal is at lower $m_{\chi}$ and smaller $\delta$. While the associated number of events predicted at XENON1T or CRESST-II in such cases are reduced, the variety of Xe and W isotopes present in either target prevents more than an ${\cal O}(1)$ suppression. As the number of events predicted at XENON1T and CRESST-II in the best fit regions to the DAMA signal for the isospin conserving case are significantly larger than what are observed, all such points remain excluded even when DM scattering off Xe or W is maximally suppressed.

\subsubsection{XENON1T} \label{sec:XE1T}

\begin{figure}
    \centering 
    \textcolor{white}{llll}\\
    \includegraphics[width=0.49\textwidth]{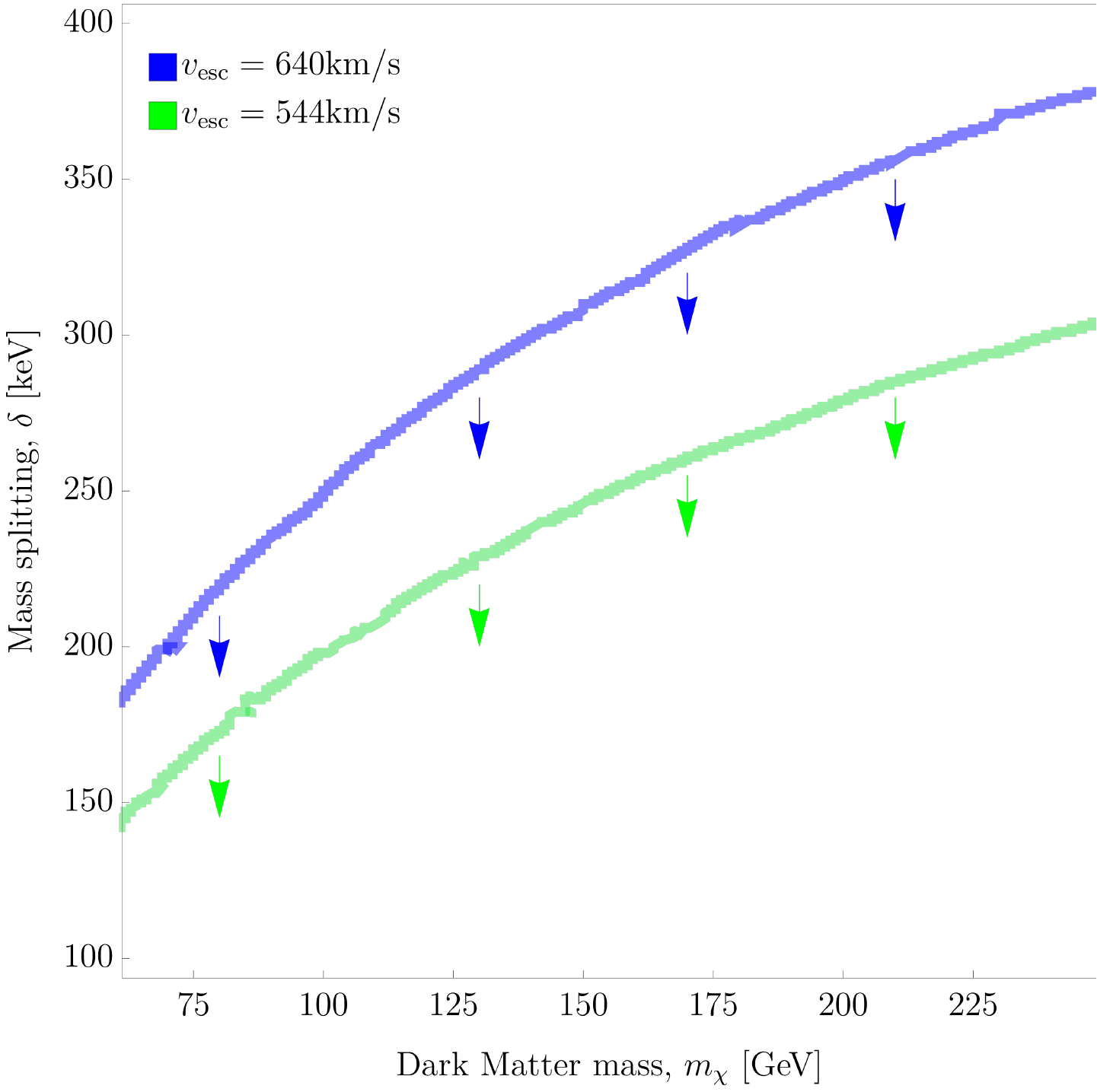}
    \includegraphics[width=0.49\textwidth]{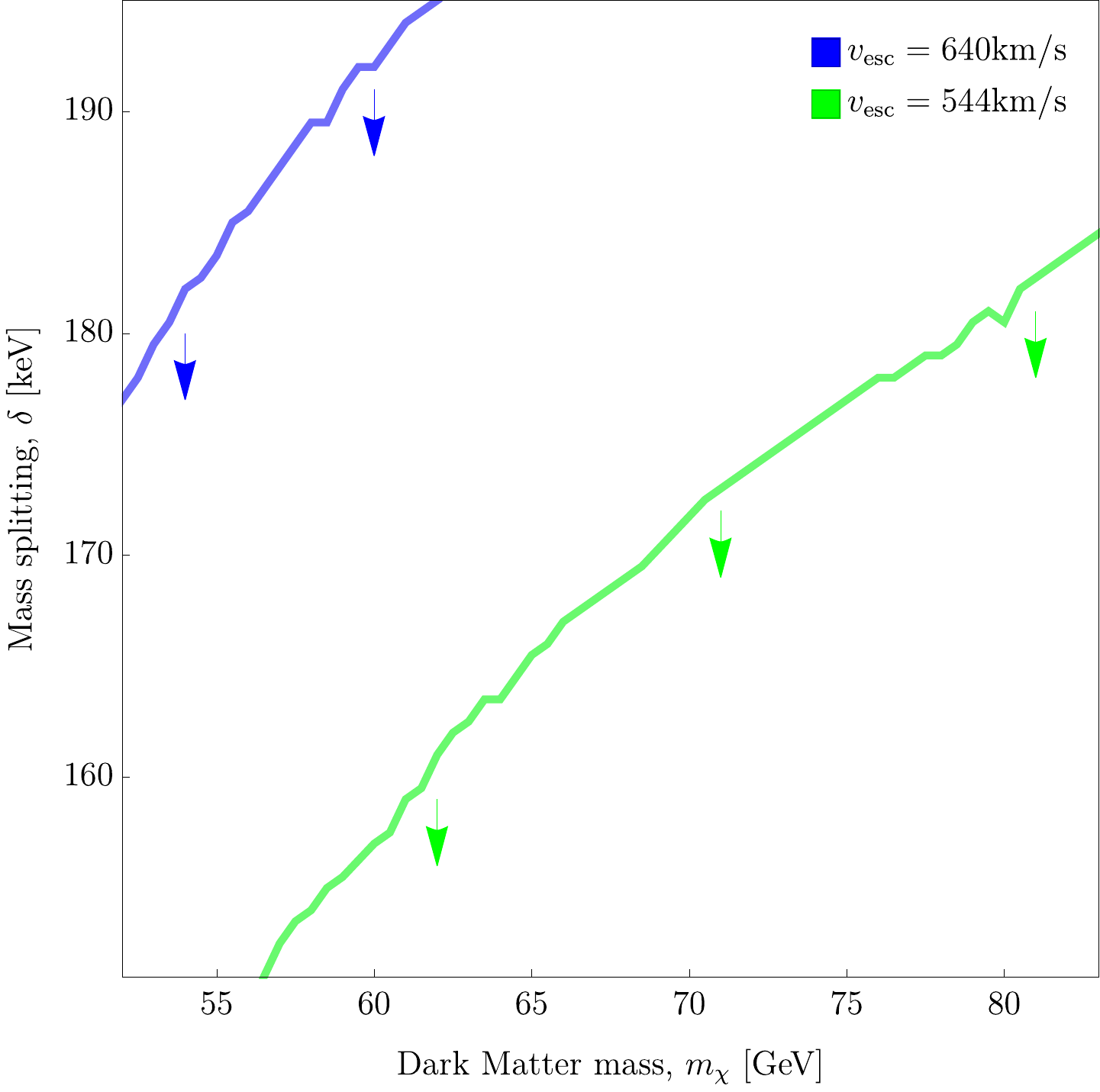}
    \caption{The regions in the $(m_{\chi}, \delta)$-plane excluded by XENON1T~\cite{Aprile:2018Xenon1TDMsearch,Aprile:2020tmw} (\emph{left}) and the remaining relevant regions excluded by CRESST-II~\cite{Angloher:2014myn} (\emph{right}) for different $v_{\rm esc}$, assuming isospin-conserving DM. The maximum velocity of the DM particles is linearly dependent on $v_{\rm esc}$. When we increase $v_{\rm esc}$, regions of parameter space previously forbidden by the threshold $v_{\rm min} < v_{\rm esc} + v_{\rm lab}(t)$ for DM-nucleus scattering become available. For the case of isospin-violating DM, the regions constrained by XENON1T and CRESST-II are approximately the same as what is presented here.}
    \label{fig: Xenon constraints for different vesc and v0}
\end{figure}

In order to constrain nuclear recoils over a larger energy range than considered in Ref.~\cite{Aprile:2018Xenon1TDMsearch}, we use data from Ref.~\cite{Aprile:2020tmw} with an accumulated exposure of 0.65$\, \rm ton\times years$. Although the latter analysis shows an excess of electron recoil events, virtually no events arising from nuclear recoils are detected. To set conservative bounds we assume that approximately $1\%$ of the $\sim 42000$ electron recoils in Ref.~\cite{Aprile:2020tmw} could be misidentified as nuclear recoils~\cite{Aprile:2019bbb}. Given a best-fit cross section for DM-Tl scattering at DAMA, the associated combination of $(m_{\chi}, \delta)$ is ruled out by XENON1T if the number of expected events corresponds to more than a $3\sigma$ statistical fluctuation in the number of misidentified electron recoils, approximately 500 events. 

Although we will show in Section~\ref{sec:BestFit} that the combined constraints from XENON1T and CRESST-II rule out all of the best-fit regions to the DAMA signal, here we focus on the points within the best-fit regions of $(m_{\chi}, \delta)$-space which are robustly constrained by XENON1T alone. When considering the large cross sections which yield the best fit to the DAMA signal in combination with the large exposure of XENON1T, such points ruled out by XENON1T alone typically predict at least 1000 events. For these points, we have checked that allowing for the cross sections to be reduced by $\lesssim 20\%$ while still keeping a $\lesssim 3 \sigma$ fit to the DAMA signal results in a prediction of at least 500 events at XENON1T and are thus still ruled out before considering the constraints from CRESST-II. Near the regions of $(m_{\chi}, \delta)$-space where DM-Xe scattering is kinematically forbidden, the expected signal in XENON1T falls off rapidly. Thus, the parts of parameter space that are not ruled out by XENON1T are close to where the kinematics of inelastic DM-Xe scattering do not allow for recoil energies of $1-201 \,$keV.

As discussed in Section~\ref{subsec: annual modulation}, the DM escape velocity and the local circular velocity are both subject to significant uncertainties (see Refs.~\cite{Ibarra:2018yxq,Wu:2019nhd} for recent analyses of astrophysical uncertainties in XENON1T constraints). The regions of parameter space that are ruled out by other direct detection experiments also depend on these parameters. In particular, the constraints from XENON1T depend heavily on the choice of astrophysical parameters since the excluded regions are determined almost entirely by kinematics. As shown in the left panel of Fig.~\ref{fig: Xenon constraints for different vesc and v0}, regions of the $(m_\chi$,$\delta)$ plane which are excluded by XENON1T grow when the escape velocity increases (there are similar effects when increasing the local circular velocity). This is a kinematical effect: The XENON1T constraints depend on the threshold for scattering, given by $v_{\rm esc} + v_{\rm lab}(t) > v_{\rm min}$. According to Eq.~\ref{eq: vmin}, the minimum velocity increases with $\delta$ and decreases with $m_{\chi}$. Thus, when $v_0$ and $v_{\rm esc}$ increase, the regions constrained by XENON1T increase in size and include larger $\delta$ and smaller $m_{\chi}$.

\subsubsection{CRESST-II} \label{sec:CR2}

We follow a similar procedure to calculate the constraints from CRESST-II.\footnote{The data from CRESST-II is more suited to our model than the data from CRESST-III~\cite{Abdelhameed_2019CRESST3} because the expected DM signal from inelastic scattering falls outside the acceptance region in CRESST-III.} We use the data from the TUM-40 detector in CRESST-II because it has the fewest observed events in the acceptance region for dark matter scattering ($E_R = 0.6-40$keV) and the highest exposure after cuts have been applied. We use the data from Ref.~\cite{Angloher:2014myn,angloher2017description} with an exposure of $ 29.35 \, \rm kg\times days$, and a cut-survival probability of $0.8$ for DM-Tungsten (W) scattering. There is an excess of events in CRESST-II for recoil energies $E_{R} \lesssim 1 \, \rm keV$, but these cannot be distinguished from the $e^{-}/\gamma$-background. To make sure no events within the $90\%$ confidence region of the $e^{-}/\gamma$-band are included, we only consider nuclear recoils with $E_{R} \gtrsim 10 \,$keV. There are no events above this limit within the signal acceptance region of the CRESST-II analysis. Given the associated best-fit cross section for DM-Tl scattering in DAMA at a point in $(m_{\chi}, \delta)$ space, we conservatively demand that the expected number of events in CRESST-II is less than 10 events. 

We find that CRESST-II rules out the remaining $(m_{\chi}, \delta)$-points in the best-fit regions, discussed in detail in Sec.~\ref{sec:BestFit}, which are not excluded by XENON1T alone. The smallest number of events predicted in CRESST-II at such points is 40 assuming the best-fit cross sections to the DAMA signal at each point. The cross sections can be reduced by $\lesssim 20\%$ while still keeping a $\lesssim 3 \sigma$ fit to the DAMA signal at each point and the smallest number of predicted events decreases to 34. Thus, the combined constraints from XENON1T and CRESST-II on the best-fit regions of $(m_{\chi}, \delta)$-space are robust to any potential reduction of the cross sections from the best-fit values while still maintaining a $\lesssim 3 \sigma$ fit to the DAMA signal.

The constraints from CRESST-II are shown in the right panel of Fig.~\ref{fig: Xenon constraints for different vesc and v0}. The regions constrained by CRESST-II behave in a similar manner to those from XENON1T when $v_{\rm esc}$ and/or $v_0$ increases. In general, the constraints from CRESST-II are less robust than those from XENON1T since CRESST-II has a much smaller exposure. However, the CRESST-II constraints are particularly important in the region of parameter space where XENON1T does not rule out the best-fit regions entirely. More specifically, when $Q_{\rm Tl} = 0.05-0.09$ XENON1T cannot entirely rule out the the associated best-fit regions due to kinematic effects. However, CRESST-II contains W target nuclei, which are more similar in mass to Tl than Xe. Thus, any best-fit regions not excluded by XENON1T are excluded by CRESST-II since W has a large enough mass for scattering with DM to occur. We also note that the regions of $(m_{\chi}, \delta)$-space not ruled out by CRESST-II are somewhat less sensitive to the kinematics of inelastic DM-W scattering and are more a function of the number of events decreasing with the best-fit cross section to the DAMA signal at lower $m_{\chi}$ and higher $\delta$.  

\section{Best-fit regions for DAMA modulation signal}
\label{sec:BestFit}

The results presented in the current section are based on the expected annual modulation signal in DAMA for $v_{\rm esc} = 544$ km/s (Mid) and $640$ km/s (High), $v_0 = 220 \,$km/s and $\rho_{\chi} = 0.3 \, \rm GeV/cm^3$. We identify the best-fit regions for two cases: isospin-conserving DM with $f_n/f_p = 1$, and isospin-violating DM with $f_n/f_p = \frac{-53}{127-53}$. For the isospin violating case, $f_n/f_p$ is chosen so that the effective coupling to I vanishes and, thus, the only non-negligible contribution in DAMA comes from DM-Tl scattering. While we allow for DM scattering off of I in the isospin-conserving case, we focus on the parameter space of the iDM model where the dominant scatting contribution comes from DM-Tl scattering. Because the scattering kinematics are very similar for I and Xe target nuclei, fits to the DAMA signal arising primarily from scattering off I are rather trivially ruled out by the XENON1T constraints described in the previous section.  The best-fit points for each value of $Q_{\rm Tl}$ are given in Table~\ref{tab:bfpoints}.

\begin{table}
\centering
\subfloat[Isospin violating DM, $v_{\rm esc} = 544 \rm km/s$]{
    \begin{tabular}{|c|cccc|} \hline
        $Q_{\rm Tl}$ & $m_{\chi}$ & $\delta$ & $\sigma_p^{\rm SI}$ & $\chi^2$  \\
        & [GeV] & [keV] & [$10^4$pb] & \\ \hline
        0.03 & 245 & 216 & $0.29$ & 35.09 \\
        0.04 & 245 & 176 & $0.11$& 12.82  \\ 
        0.05 & 87.5 & 170  & $0.61$  & 17.45 \\
        0.06 & 75 & 168 &  $2.02$ & 19.05 \\
        0.07 & 70 & 164 &  $3.03$ & 18.55 \\
        0.08 & 70 & 164 & $3.02$ & 18.54  \\ 
        0.09 & 70 & 160 & $1.51$ & 21.99 \\ \hline
        \end{tabular}}
\subfloat[Isospin violating DM, $v_{\rm esc} = 640 \rm km/s$]{
    \begin{tabular}{|c|cccc|} \hline
        $Q_{\rm Tl}$ & $m_{\chi}$ & $\delta$ & $\sigma_p^{\rm SI}$ & $\chi^2$  \\
        & [GeV] & [keV] & [$10^4$pb] & \\ \hline
        0.03 & 245 & 248 & 0.69 & 27.80 \\
        0.04 & 247.5 & 178 & $0.10$& 12.69  \\ 
        0.05 & 72.5 & 190  & $3.69$  & 17.27 \\
        0.06 & 65 & 188 &  $9.49$ & 19.03 \\
        0.07 & 60 & 184 &  $20.7$ & 18.68 \\
        0.08 & 60 & 184 & $20.4$ & 18.52  \\ 
        0.09 & 60 & 178 & $7.35$ & 21.77 \\ \hline
        \end{tabular}}
\hspace{0mm}
\subfloat[Isospin conserving DM, $v_{\rm esc} = 544 \rm km/s$]{
            \begin{tabular}{|c|cccc|} \hline
        $Q_{\rm Tl}$ & $m_{\chi}$ & $\delta$ & $\sigma_p^{\rm SI}$ & $\chi^2$  \\
        & [GeV] & [keV] & [pb] & \\ \hline
        0.03 & 82.5 & 174 & $14.9$ & 61.65 \\
        0.04 & 85 & 177 & $14.5$& 22.57  \\ 
        0.05 & 78.5 & 169.25  & $17.4$  & 12.67 \\
        0.06 & 74.75 & 167 &  $24.5$ & 18.68 \\
        0.07 & 71 & 165 &  $39.8$ & 17.32 \\
        0.08 & 69.25 & 163 & $42.5$ & 17.63  \\ 
        0.09 & 69 & 160 & $24.6$ & 21.74 \\ \hline
        \end{tabular}}
\subfloat[Isospin conserving DM, $v_{\rm esc} = 640 \rm km/s$]{
    \begin{tabular}{|c|cccc|} \hline
        $Q_{\rm Tl}$ & $m_{\chi}$ & $\delta$ & $\sigma_p^{\rm SI}$ & $\chi^2$  \\
        & [GeV] & [keV] & [pb] & \\ \hline
        0.03 & 100 & 236 & $56.7$ & 59.77 \\
        0.04 & 78 & 208 & $119$& 34.67  \\ 
        0.05 & 65.75 & 190  & $134$  & 12.15 \\
        0.06 & 63.75 & 186.5 &  $385$ & 17.93 \\
        0.07 & 61 & 185.25 &  $243$ & 17.30 \\
        0.08 & 59.5 & 182.75 & $254$ & 17.64  \\
        0.09 & 59.25 & 177.5 & $105$& 21.18\\ \hline
        \end{tabular}}
    \caption{The best fit points for isospin violating (\emph{top}) and isospin conserving (\emph{bottom}) scattering in DAMA, assuming either $v_{\rm esc} = 544 \, \rm km/s$ (\emph{left}) or $v_{\rm esc} = 640 \,$km/s (\emph{right}). We have fixed $\rho_{\chi} = 0.3 \, \rm GeV/cm^3$ and $ v_{0} = 220 \, \rm km/s$ in all cases. Note that when $\delta < 140 \, $keV in the isospin conserving case, there exist points with better fits to the DAMA data that arise from DM-I scattering. However, we focus on cases where the DAMA signal arises from DM-Tl scattering.}
    \label{tab:bfpoints}
\end{table}

\begin{figure}
    \centering
    \includegraphics[width=\textwidth]{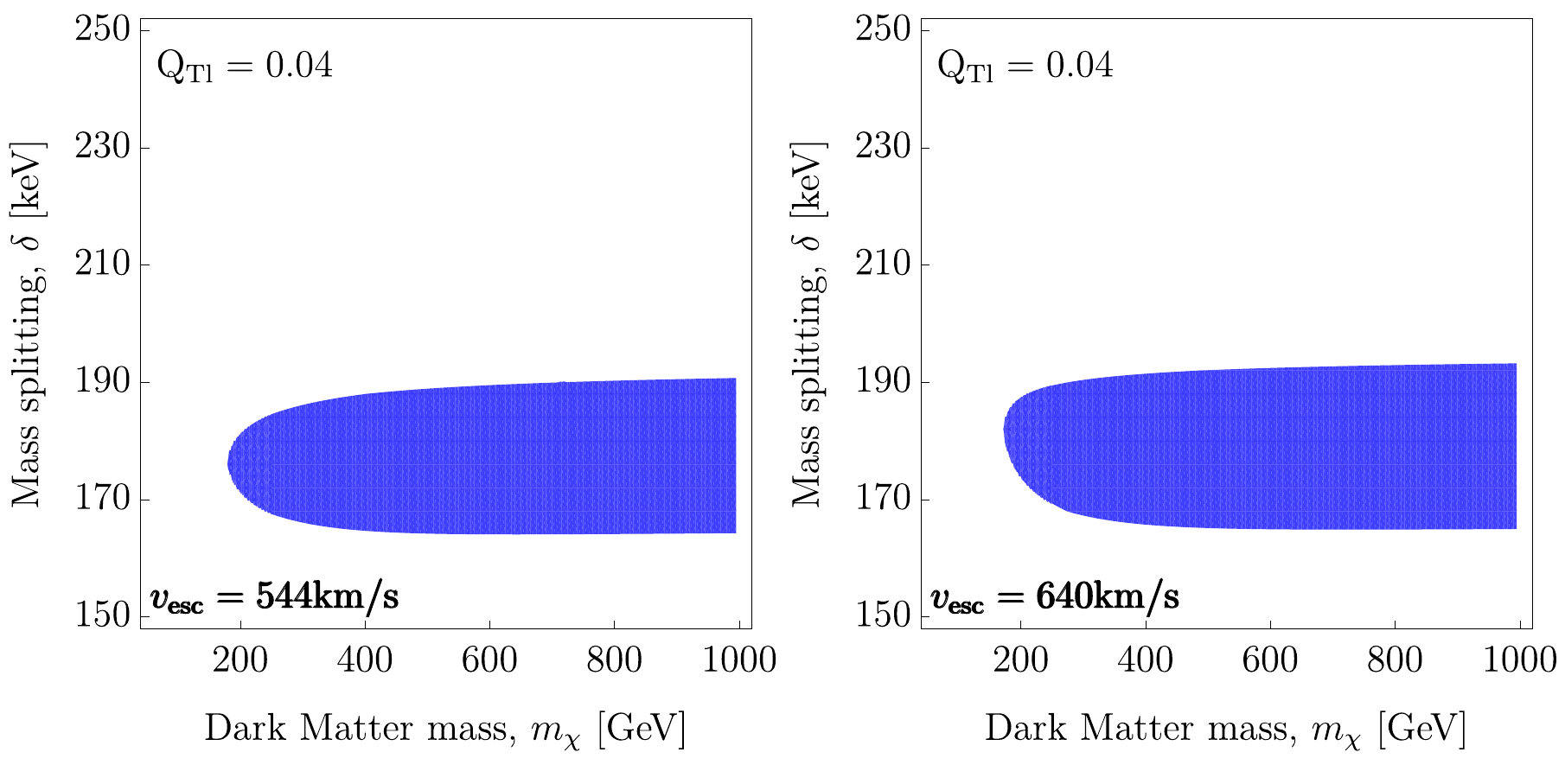}
    \caption{The best-fit regions within $2\sigma$ for DM-Tl scattering where DM-I scattering is maximally suppressed by isospin violation, as described in Eq.~\eqref{eq: isospin violating case}. \emph{Left}: The best-fit region assuming the astrophysical parameters $\rho_{\chi} = 0.3 \, \rm GeV/cm^3$, $ v_{0} = 220 \, \rm km/s$ and $ v_{\rm esc} = 544 \, \rm km/s$. \emph{Right}: The best-fit region assuming $v_{\rm esc} = 640 \,$km/s and keeping all other astrophysical parameters the same. In both cases, there are no substantial regions of parameter space within $2\sigma$ for Tl quenching factors other than $Q_{\rm Tl} = 0.04$. The best-fit regions extend indefinitely towards larger $m_{\chi}$, with the  corresponding range of $\delta$ remaining constant. The shaded regions in both panels, corresponding to the only substantial iDM parameter space where DM-Tl scattering fits the DAMA signal within $2\sigma$, are ruled out by the constraints from XENON1T shown in Fig.~\ref{fig: Xenon constraints for different vesc and v0}.}
    \label{fig: 2sigma best-fit regions, isospin-violating}
\end{figure}

\begin{figure}
\centering
    \includegraphics[width=\textwidth]{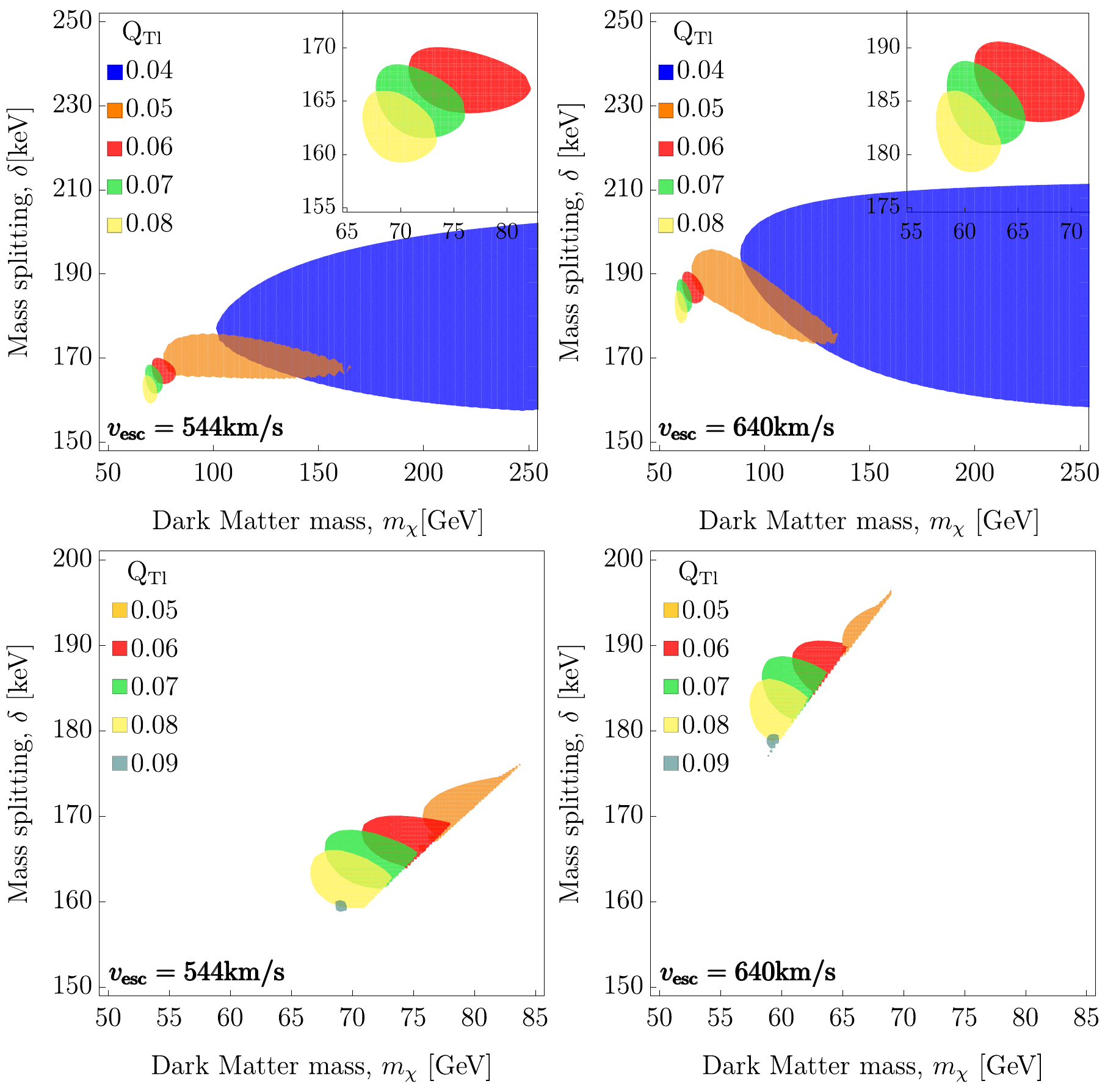}
    \caption{
  The best-fit regions within $3\sigma$ for isospin violating (\emph{top}) and isospin conserving (\emph{bottom}) scattering in DAMA, assuming either $v_{\rm esc} = 544 \,$km/s (\emph{left}) or $v_{\rm esc} = 640 \,$km/s (\emph{right}). Note that the $3\sigma$ regions for $ Q_{\rm Tl} = 0.04$ extend indefinitely towards higher $m_{\chi}$ in the same manner as the $2\sigma$ regions presented in Fig.~\ref{fig: 2sigma best-fit regions, isospin-violating}, with the same range of $\delta$ at larger $m_{\chi}$. Note there are no best-fit regions within $3\sigma$ when $Q_{\rm Tl} = 0.02$ and $0.03$. For the isospin conserving case, the regions within $3\sigma$ corresponding to different Tl quenching factors come from DM-Tl scattering only since the minimum velocities required to scatter in these parts of parameter space are too large for DM-I scattering to occur. This is illustrated by the sharp cutoff starting at $m_{\chi} = 70 \, \rm GeV$ and $\delta \sim 160 \, \rm GeV$ in the lower left panel (starting at $m_{\chi} \sim 60 \,$GeV and $\delta \sim 175 \,$keV in the lower right panel) which shows where the DM-I scattering becomes kinematically forbidden. All regions shown in this figure are ruled out by a combination of the XENON1T and CRESST-II constraints shown in Fig.~\ref{fig: Xenon constraints for different vesc and v0}.}
    \label{fig: BF regions 3sigma}
\end{figure}

In Figure~\ref{fig: 2sigma best-fit regions, isospin-violating} we show the only interesting $2\sigma$ best-fit region within the various scenarios we have explored, the isospin-violating case with $Q_{\rm Tl} = 0.04$. For any of the other Tl quenching factors we consider here, there are no substantial regions of parameter space in which the dominant scattering contribution comes from DM-Tl scattering that fit the DAMA annual modulation signal at $2\sigma$.\footnote{There is a $2\sigma$ region in the isospin-conserving case ranging from $\delta = 100-120 \,$keV and $m_{\chi} = 50-250 \,$GeV that comes from DM-I scattering. This region is completely ruled out by XENON1T.} When $ Q_{\rm Tl} = 0.05$, there is a tiny $2\sigma$ region in the isospin-conserving case. This region is a thin line centered near $m_\chi \simeq 80 \,$GeV and $\delta \simeq 171 \,$keV, where the parameter space is ruled out by CRESST-II.\footnote{The thin line lies along the straight edge of the $3\sigma$ best-fit contour for $ Q_{\rm Tl} = 0.05$ is shown in the bottom left panel of Fig.~\ref{fig: BF regions 3sigma}.} For both escape velocities considered in our analysis, we see that the $2\sigma$ best-fit regions in the isospin-violating case with $Q_{\rm Tl} = 0.04$ have $165 \, {\rm keV} \lesssim \delta \lesssim 190 \, {\rm keV}$ (up to $195 \, \rm keV$ when $v_{\rm esc} = 640 \, \rm km/s$) and extend indefinitely towards higher $m_{\chi}$. As $m_{\chi}$ increases, the expected DM-Tl scattering signal reaches a plateau and no longer changes with increasing DM mass. However, the $2\sigma$ best-fit regions for $Q_{\rm Tl} = 0.04$ and both escape velocities are completely ruled out by XENON1T, as the suppression of scattering off of Xe targets becomes less effective for larger DM masses.

The best-fit regions within $3\sigma$ are presented in Fig.~\ref{fig: BF regions 3sigma}. The $3\sigma$ best-fit regions for the isospin-violating case with $\rm Q_{Tl} = 0.04$ extend to significantly lower DM masses, $m_\chi \gtrsim 100 \, {\rm GeV}$, than the corresponding $2\sigma$ best-fit regions. Also, at even lower DM masses, there are several smaller $3\sigma$ best-fit regions with $\rm Q_{Tl} > 0.04$. In isospin-conserving cases, we see that the $3\sigma$ best-fit regions vanish for smaller quenching factors because DM-I scattering becomes kinematically allowed at higher DM masses but does not yield a good fit to the DAMA modulation signal. 

Comparing Figs.~\ref{fig: Xenon constraints for different vesc and v0} and~\ref{fig: BF regions 3sigma}, we see that XENON1T rules out all of the $3\sigma$ best-fit regions for $m_\chi \gtrsim 100 \, {\rm GeV}$ and part of the $3\sigma$ best-fit regions for $m_\chi \lesssim 100 \, {\rm GeV}$, while CRESST-II rules out all lower mass $3\sigma$ best-fit regions. Since CRESST-II contains W, which has mass nearly as large as Tl, DM-W scattering is kinematically allowed for approximately all mass and $\delta$ ranges where DM-Tl scattering occurs. Thus, the parts of the $3\sigma$ best-fit regions not excluded by XENON1T are excluded by CRESST-II. We therefore find that a combination of XENON and CRESST data rules out all regions of parameter space for which inelastic scattering could explain the DAMA data.

\section{Conclusions}
\label{sec:Conclusions}
In this paper we have studied the compatibility of the observed DAMA modulation signal with inelastic dark matter (DM) scattering in DAMA. This type of scattering has been proposed as a solution to the discrepancy between the observed signal in DAMA and the lack of signal in other direct detection experiments. In particular, it has been proposed that the DM primarily scatters off of the $0.1\%$ Thallium (Tl) dopant in the DAMA crystals. Since inelastic scattering prefers heavy nuclei, it is possible for DM scattering off of atoms much lighter than Tl, such as xenon, to be kinematically forbidden. Thus, if inelastic scattering gives the leading contribution to the DM-nucleon scattering cross-section, the constraints from liquid noble gas target experiments, such as XENON1T~\cite{Aprile:2017Xenon1T,Aprile:2018Xenon1TDMsearch}, can be suppressed. Previously, Chang, Lang and Weiner~\cite{Chang:2010pr} in 2010 and more recently the DAMA collaboration~\cite{Bernabei_2019Modeldependentanalysis} in 2019 found regions in parameter space where DAMA data was consistent with an interpretation of inelastic scattering of DM with Tl in the detector.  In light of the recent DAMA results, we have expanded upon these previous analyses by using the most recent data from DAMA/LIBRA phase I and II~\cite{Bernabei:2013phase1, Bernabei:2018phase2} and including constraints from both XENON1T~\cite{Aprile:2018Xenon1TDMsearch,Aprile:2020tmw} and CRESST-II~\cite{Angloher:2014myn}. 

We have compared the expected modulation signal for inelastic DM scattering in DAMA with the observed signal for two cases: Isospin-conserving DM and isospin-violating DM, where DM-I scattering is maximally suppressed. The main sources of uncertainty are the quenching factor of Tl and the astrophysical parameters $v_0, v_{\rm esc}$, and $\rho_\chi$. The quenching factor of Tl has never been measured directly, but an approximate range was found by Chang et al. in their 2010 paper. As a result of their calculations, we have performed our analysis for a range of Tl quenching factors from $0.03$ to $0.09$ with a step size of $0.01$. We have shown that the best-fit regions depend heavily on the choice of quenching factor, since the quenching factor directly affects the shape of the modulation signal (see Fig.~\ref{fig: form factor and energy spectrum, different QFs}). We have also considered how the fits to the DAMA signal depend on the astrophysical parameters $v_{\rm esc}$ and $v_0$. We have found that the best-fit regions increase slightly in size and are shifted towards higher mass splitting and lower DM masses when these two parameters are increased.

We find that all regions in $(m_{\chi}, \delta)$ space that give a good fit to the DAMA modulation signal are ruled out by XENON1T, CRESST-II, or both. For the isospin-violating case there is a $2\sigma$ region when $Q_{\rm Tl} = 0.04$. For the isospin-conserving case, we have found a $2\sigma$ region for $\delta < 120 \,$keV that comes from DM-I scattering. Constraints from XENON1T exclude all of the $2\sigma$ best-fit regions.  The $3\sigma$ best-fit regions are ruled out by XENON1T (high mass) and/or CRESST-II (low mass). The constraints from XENON1T and CRESST-II have been found by directly comparing the number of events that have been observed in these experiments with what is expected given the Tl interpretation of the DAMA signal. The best-fit regions presented in the paper by the DAMA collaboration~\cite{Bernabei_2019Modeldependentanalysis} are also within the parameter space that is ruled out by XENON1T and/or CRESST-II. Based on these results, we conclude that inelastic DM-Tl scattering in DAMA cannot explain the discrepancy between the modulation signal in DAMA and the null results of other experiments. 

\acknowledgments
The authors would like to thank S.~Baum, T.~Edwards and J.~Kumar for useful discussions. S.~Jacobsen, K.~Freese and P.~Stengel acknowledge support by the Vetenskapsr\r{a}det (Swedish Research Council) through contract No. 638-2013-8993 and the Oskar Klein Centre for Cosmoparticle Physics. K. Freese is the Jeff \& Gail Kodosky Endowed Chair of Physics at University of Texas in Austin and is grateful for support. K.~Freese acknowledges support from DoE grant DE-SC007859 at the University of Michigan. The work of P.~Sandick is supported in part by NSF grant PHY-1720282. The work of P.~Stengel is partially supported by the research grant  ``The Dark Universe: A Synergic Multi-messenger Approach'' number 2017X7X85K under the program PRIN 2017 funded by the Ministero dell'Istruzione, Universit{\`a} e della Ricerca (MIUR), and by the ``Hidden'' European ITN project  (H2020-MSCA-ITN-2019//860881-HIDDeN). 

\printbibliography
\end{document}